\documentclass[aps,prb,floatfix,superscriptaddress,reprint]{revtex4-1} 
\usepackage{graphicx}
\usepackage{amsmath,amsfonts,amssymb}
\usepackage{color}
\usepackage{pdfpages}

\makeatletter
\AtBeginDocument{\let\LS@rot\@undefined}
\makeatother
     
\begin{document}

\title{Critical Spin Liquid versus Valence Bond Glass in Triangular Lattice Organic $\kappa$-(ET)$_2$Cu$_2$(CN)$_3$}
\date{\today}
\begin{abstract}
In the quest for materials with unconventional quantum phases,
the organic triangular-lattice antiferromagnet 
$\kappa$-(ET)$_2$Cu$_2$(CN)$_3$ has been extensively discussed as a quantum spin liquid (QSL) candidate.
Recently, an intriguing quantum critical behaviour was suggested
from low-temperature magnetic torque experiments.
Through microscopic analysis of all anisotropic contributions, including Dzyaloshinskii-Moriya and multi-spin scalar chiral interactions, we highlight significant deviations of the experimental observations
from a quantum critical scenario.
Instead, we show that disorder-induced spin defects provide a comprehensive explanation of the
 low-temperature properties. These spins are attributed to valence bond defects that emerge spontaneously as the QSL enters
a valence bond glass phase at low temperature. This theoretical
treatment
is applicable to a general class of frustrated magnetic systems
and has important implications for the interpretation of 
magnetic torque, nuclear magnetic resonance, thermal transport and
thermodynamic experiments.
\end{abstract}

\author{Kira Riedl}
\affiliation{Institut f\"ur Theoretische Physik, Goethe-Universit\"at Frankfurt,
Max-von-Laue-Strasse 1, 60438 Frankfurt am Main, Germany}
\author{Roser Valent{\'\i}}
\affiliation{Institut f\"ur Theoretische Physik, Goethe-Universit\"at Frankfurt,
Max-von-Laue-Strasse 1, 60438 Frankfurt am Main, Germany}
\author{Stephen M. Winter*}
\affiliation{Institut f\"ur Theoretische Physik, Goethe-Universit\"at Frankfurt,
Max-von-Laue-Strasse 1, 60438 Frankfurt am Main, Germany}

\maketitle


Despite intensive efforts devoted to uncover
the nature of the low-temperature
properties of the QSL candidate
$\kappa$-(ET)$_2$Cu$_2$(CN)$_3$
($\kappa$-Cu)\cite{shimizu2003spin,kanoda2011mott,zhou2017quantum}, a few puzzles remain unresolved, such as
the presence of an anomaly near $T^* =$ 6 K in a wide variety of
experiments including $^{13}$C nuclear magnetic resonance
(NMR)\cite{shimizu2006emergence}, muon spin resonance
($\mu$SR)\cite{pratt2011magnetic,nakajima2012microscopic}, electron spin
resonance (ESR)\cite{padmalekha2015esr}, specific
heat\cite{yamashita2008thermodynamic,manna2010lattice}, ultrasound
attenuation\cite{poirier2014ultrasonic}, and thermal
expansion\cite{manna2010lattice}. Various scenarios have been suggested to
explain this anomaly such as spin-chirality
ordering\cite{baskaran1989novel}, a spinon-pairing
transition\cite{PhysRevLett.94.127003,grover2010weak,PhysRevLett.98.067006,PhysRevLett.99.266403}
or formation of an exciton condensate\cite{qi2008insulator}.  
However, a
comprehensive explanation of all aspects of the anomaly
and its possible relation to a spin liquid phase is currently lacking.

\begin{figure} \includegraphics[width=\columnwidth]{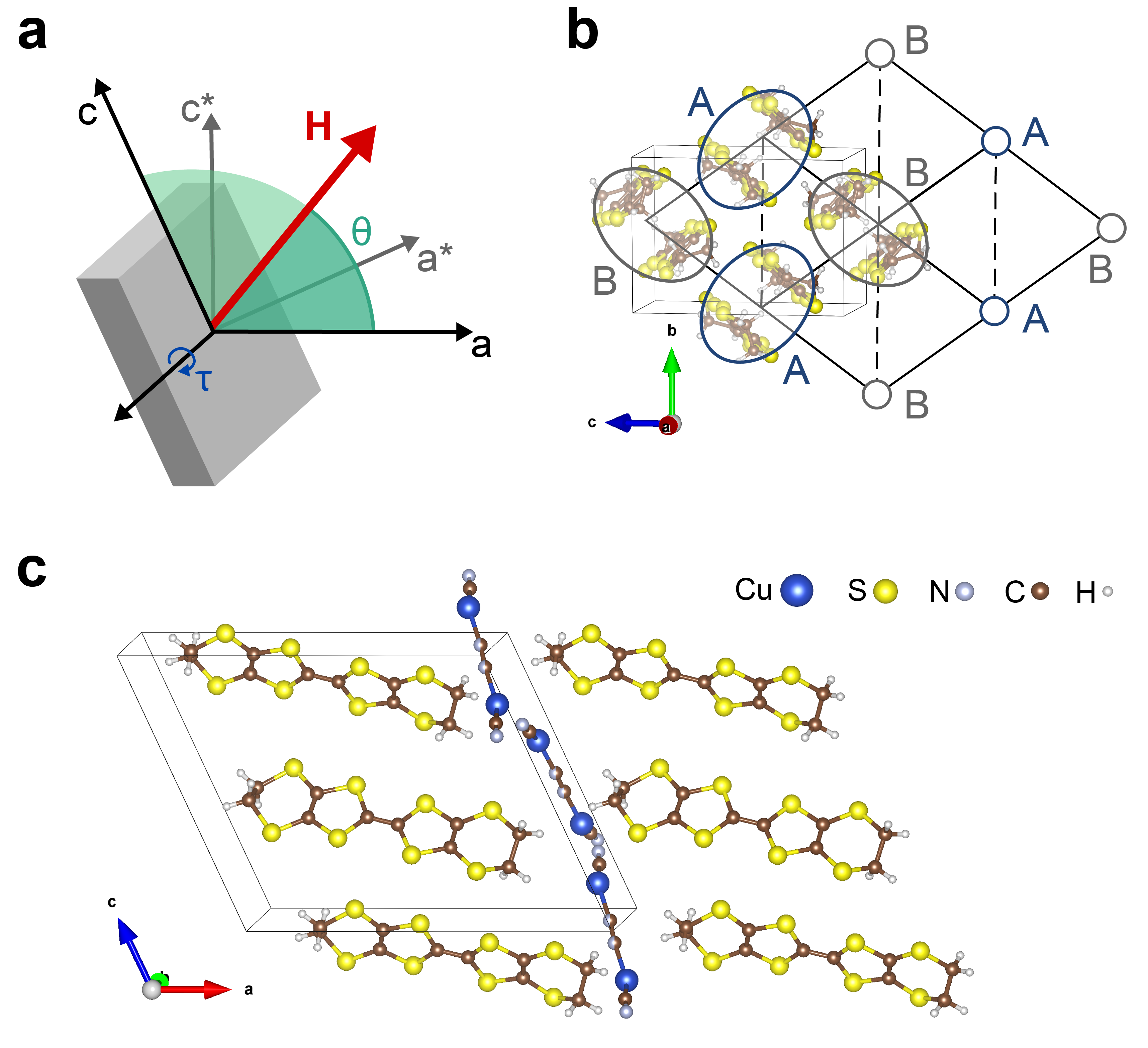}
\caption{\textbf{Magnetic torque in the organic charge transfer salt
$\kappa$-Cu.} (a) Definition of quantities used in the torque expressions.
$\alpha$ is the angle between the $a^\ast$ axis and the principle axis of
$(\mathbb{G}_u \cdot \mathbb{G}_u^{\rm T})$, which is approximately the $a$
axis.  $\theta$  is the angle between
the magnetic field $\mathbf{H}$ and 
the principle axis of
$(\mathbb{G}_u \cdot \mathbb{G}_u^{\rm T})$.
(b) Anisotropic triangular lattice with the sublattices labelled A and B on top
of the ET dimers in $\kappa$-Cu. (c) Crystal structure of $\kappa$-Cu, showing
the organic layer of ET molecules and the anion layer.} \label{fig:setup}
\end{figure}

In this work, we show that significant insight can be obtained from a microscopic analysis of
 magnetic torque measurements~\cite{isono2016quantum}. A finite torque $\tau$ reflects a variation of the
energy $E$ as a function of
the orientation of the external magnetic field $\mathbf{H}$ (Fig.~\ref{fig:setup}a):
 \begin{align} \label{eq:torque_gen}
\tau = \frac{\text{d} E}{\text{d} \theta}
\end{align}
where $\theta$ is the angle between $\mathbf{H}$ and a reference axis. Conventionally, $\tau$ is considered to probe the uniform magnetization $\mathbf{M}$, through $\vec{\tau} \propto \mathbf{M}\times \mathbf{H}$. However, the torque is sensitive to {\it any} angular variation of the energy, which we show, in this work, allows more general instabilities to be probed. We demonstrate this through consideration of the recently reported torque response of $\kappa$-Cu~\cite{isono2016quantum}, which 
displays several characteristic
features. 

For $\kappa$-Cu, as $\mathbf{H}$ is rotated in the $ac^*$-plane, $\tau$ follows a sinusoidal angle dependence $\tau \propto \sin2 (\theta-\theta_0)$, with an angle shift $\theta_0$ that increases substantially at low $T$ and $H$, suggesting the emergence of a new contribution below $T^*$. Indeed, the torque susceptibility
$\chi_\tau = \tau/H^2$ diverges as a power law $\chi_\tau \sim T^{-\omega}$ at low-fields, and as $\chi_\tau \sim
H^{-\zeta}$ at low temperature, with approximately
the same exponent~\cite{isono2016quantum} $\omega \approx \zeta \approx 0.8$. Consequently the torque
displays an apparent critical $H/T$ scaling over several orders of magnitude. In this context, the remarkable behaviour of $\kappa$-Cu has been interpreted\cite{isono2016quantum} in
terms of a field-induced quantum critical point with a diverging uniform spin susceptibility. However, the presence of magnetic interactions beyond the conventional Heisenberg couplings allows more general instabilities of the spin liquid to be directly probed by torque measurements. 
For example, in $\kappa$-Cu, spin-orbit coupling (SOC) leads to a
staggered $g$-tensor and a finite Dzyaloshinskii-Moriya (DM)
interaction\cite{winter2017SOC}. These terms ensure torque contributions from the staggered magnetic susceptibility, which may diverge near an instability towards a {\it
canted} N\'eel antiferromagnetic order\cite{kagawa2008field}. Furthermore, higher order 
 multi-spin interactions\cite{motrunich2005variational} couple the field to the scalar spin chirality, allowing $\tau$ to probe transitions to chiral phases\cite{baskaran1989novel}.

Importantly, through a microscopic
calculation of $\tau$, we first show that the quantum critical scenario cannot account for the experimental findings, even when all such instabilities are considered. Instead, motivated by the observation of an inhomogeneous NMR response below $T^*$, we consider
 disorder-induced effects~\cite{kimchi2017valence,kimchi2018heat}. In particular, we show that the NMR and $\tau$ experiments are consistent with proximity to a finite randomness large-spin fixed point~\cite{westerberg1995random,westerberg1997low}, where the diverging $\chi_\tau$ reflects a small density of quasi-free defect spins. By considering the specific temperature dependence of the response, the $T^*$ anomaly can be interpreted as the onset of a valence bond glass, at which the resonating valence bonds of the QSL become randomly pinned. Finally, we discuss how this interpretation is consistent with a remarkable range of experiments.

\section{Results}

\textbf{Effective Hamiltonian.}
We first consider possible anisotropic terms in the spin Hamiltonian that contribute to $\tau$. These include anisotropic exchange interactions and $g$-tensor anisotropy that arise from spin-orbit coupling (SOC), as well as higher order spin-chiral terms. {\it Ab-initio} estimates of each contribution are detailed in Supplementary Note 1.

The bilinear exchange interactions can be written in terms of the Heisenberg exchange $J_{ij}$, the Dzyaloshinskii-Moriya (DM) vector $\mathbf{D}_{ij}$, and the pseudo-dipolar tensor $\Gamma_{ij}$:
\begin{align} \label{eq:Hamiltonian(2)}
\mathcal{H}_{ij}=J_{ij} \ \mathbf{S}_i \cdot \mathbf{S}_j + \mathbf{D}_{ij} \cdot ( \mathbf{S}_i \times \mathbf{S}_j ) + \mathbf{S}_i \cdot \Gamma_{ij} \cdot \mathbf{S}_j
\end{align} 
where $\mathbf{S}_i$ denotes the spin at site $i$. For $\kappa$-Cu, the sites consist of molecular dimers, which are arranged on an anisotropic triangular lattice with two dimer sublattices, labelled A and B in Fig.~\ref{fig:setup}b. We have previously estimated the bilinear interactions including SOC for $\kappa$-Cu in Ref.~\onlinecite{winter2017SOC}, and found that the DM-vectors are nearly parallel, but staggered ($\mathbf{D}_{ij} \approx \pm \mathbf{D}$), with magnitude $|\mathbf{D}|/J \sim 5\%$. It is therefore useful to transform the Hamiltonian\cite{shekhtman1992staggered} by applying local staggered
rotations of the spins $\mathbf{S}_i \rightarrow \tilde{\mathbf{S}}_i$ around
$\mathbf{D}$ by a canting angle $\pm \phi_i$. For a purely staggered $\mathbf{D}$-vector, this eliminates the DM interaction and also,
in leading orders, the pseudo-dipolar tensor. As a result, the transformed bilinear interactions are:
\begin{align}\label{eqn3}
\mathcal{H}_{ij} \approx \tilde{J}_{ij} \ \tilde{\mathbf{S}}_i \cdot \tilde{\mathbf{S}}_j 
\end{align}
with $\tilde{J}_{ij} \approx J_{ij}$ and $J_{ij}/k_B \sim$ 230 - 270 K. Since the transformed bilinear interactions are rotationally invariant, they do not explicitly contribute to $\tau$. 
Instead, the anisotropic effects are shifted to the transformed Zeeman term: 
\begin{align}
\mathcal{H}_{\rm Zee, eff} =-\mu_B \mathbf{H} \cdot \sum_i ( \tilde{\mathbb{G}}_u + \eta_i \tilde{\mathbb{G}}_s ) \cdot \tilde{\mathbf{S}}_i,
\end{align}
where $\tilde{\mathbb{G}}_u$ and $\tilde{\mathbb{G}}_s$ are the uniform and staggered components of the $g$-tensor in the rotated frame, and $\eta_i = +1 (-1)$ for $i \in \text{A} (\text{B})$ sublattice. Microscopic expressions for $\tilde{\mathbb{G}}_u$ and $\tilde{\mathbb{G}}_s$ are detailed in Supplementary Note 1.

For materials close to the Mott transition (such as $\kappa$-Cu), it has been
suggested that higher order ring-exchange terms such as $\mathcal{H}_{(4)}=1/S^2\sum_{\langle ijkl \rangle} \tilde{K}_{ijkl} (\tilde{\mathbf{S}}_i\cdot
\tilde{\mathbf{S}}_j)(\tilde{\mathbf{S}}_k\cdot\tilde{\mathbf{S}}_l)$
may also play a significant role in stabilizing QSL states\cite{holt2014spin,motrunich2005variational,block2011spin,holt2014spin}.
As detailed in Supplementary Note 1, our {\it ab-initio} estimates find $\tilde{K}/\tilde{J} \sim 0.1$, which represents a significant contribution~\cite{motrunich2005variational,holt2014spin}. Including SOC leads to anisotropic ring-exchange terms, in principle. 
Fortunately, the leading order contributions to these terms are also eliminated by the $\mathbf{S}\to \tilde{\mathbf{S}}$ transformation, such that the transformed 4-spin ring exchange terms do not explicitly contribute to $\tau$ at lowest order.

Finally, additional 3-spin scalar chiral interactions\cite{motrunich2005variational} may also arise when a finite magnetic flux penetrates the 2D organic layers. In the rotated coordinates,
this provides the interaction:
\begin{align}
\mathcal{H}_{\Phi,\rm eff} = -\mu_B  j_\Phi \, (\mathbf{H} \cdot \mathbf{n})  \sum_{\langle ijk \rangle} \tilde{\mathbf{S}}_i\cdot(\tilde{\mathbf{S}}_j \times \tilde{\mathbf{S}}_k ),
\end{align}
where $\mathbf{n}$ is the out-of-plane unit vector and the parameter $j_\Phi$ is proportional to the magnetic flux $\frac{\hbar}{q} \Phi$ enclosed by the triangular plaquette $\langle ijk \rangle$. Since only the out-of-plane component of the magnetic field couples to the scalar spin chirality, this term contributes explicitly to $\tau$. We estimate $\mu_B j_\Phi/k_B \sim $ 0.03 K/T.
\\


\begin{figure*}
\includegraphics[width=\textwidth]{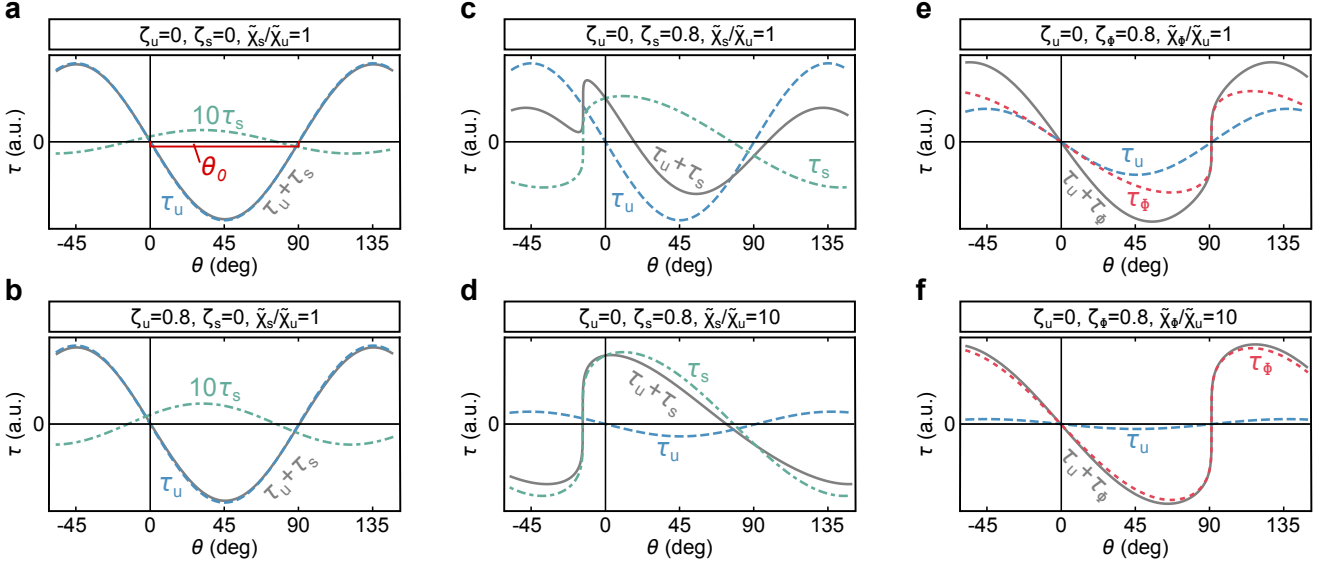}
\caption{\textbf{Angle-dependence of bulk torque for indicated parameter sets.} The total bulk torque is illustrated with solid gray, the uniform with dashed light blue, the staggered with dotted dashed blue, and the chiral contribution with dotted green lines. (a) Conventional case with field-independent susceptibility, (b) Diverging uniform susceptibility ($\zeta_u=0.8$), (c) Diverging staggered susceptibility ($\zeta_s=0.8$) at intermediate field ($\tilde{\chi}_s/\tilde{\chi}_u=1$), (d) Diverging staggered susceptibility ($\zeta_s=0.8$) at low-field ($\tilde{\chi}_s/\tilde{\chi}_u=10$), (e) Diverging scalar chiral susceptibility ($\zeta_\Phi=0.8$) at intermediate field ($\tilde{\chi}_\Phi/\tilde{\chi}_u=1$), (f) Diverging scalar chiral susceptibility ($\zeta_\Phi=0.8$) at low-field ($\tilde{\chi}_\Phi/\tilde{\chi}_u=10$). } \label{fig:staggered}
\end{figure*}

\textbf{Bulk torque expressions near a QCP.} In the rotated frame, the bulk contribution to $\tau$ depends only on two terms in the Hamiltonian:
\begin{align} \label{eq:torque_pre}
\tau 
&= \frac{\rm{d}\langle \mathcal{H}_{\rm Zee,eff} \rangle}{\rm{d} \theta} + \frac{\rm{d} \langle \mathcal{H}_{\rm \Phi, eff} \rangle}{\rm{d} \theta}.
\end{align}  
 For notational convenience, we introduce effective fields, $\mathbf{H}_{\text{eff},u/s}=\tilde{\mathbb{G}}_{u/s}^{\text{T}} \cdot \mathbf{H}$ and $\mathbf{H}_{\text{eff},\Phi}=j_\Phi (\mathbf{n} \cdot \mathbf{H}) \mathbf{n}$, which couple directly to the rotated spin variables:
\begin{align}
\mathcal{H}_{\rm Zee, eff} =-\mu_B \sum_i ( \mathbf{H}_{\text{eff},u} + \eta_i \mathbf{H}_{\text{eff},s} ) \cdot \tilde{\mathbf{S}}_i, \\
\mathcal{H}_{\Phi,\rm eff} = -\mu_B \vert \mathbf{H}_{\text{eff},\Phi} \vert \sum_{\langle ijk \rangle} \tilde{\mathbf{S}}_i\cdot(\tilde{\mathbf{S}}_j \times \tilde{\mathbf{S}}_k ).
\end{align}
Since the transformed Hamiltonian of Eq.~\eqref{eqn3} is isotropic, the energy is minimized when the spin expectation values $\langle \sum_i \tilde{\mathbf{S}}_i \rangle$ and $\langle \sum_i \eta_i \tilde{\mathbf{S}}_i \rangle$ are parallel to the corresponding effective fields $\mathbf{H}_{\text{eff},u}$ and $\mathbf{H}_{\text{eff},s}$, respectively. However, in order to compare with experimental results, it is more convenient to express $\tau$ in terms of the magnitude $H$ and orientation $\theta$ of the original laboratory field $\mathbf{H}$. Evaluating Eq.~\eqref{eq:torque_pre} shows that the $H$ and $\theta$ dependences are separable (see Supplementary Note 2 for detailed derivation). The total bulk magnetic torque is the sum of the uniform, the staggered, and the chiral contributions $\tau_{\rm B}=\tau_u+\tau_s+\tau_\Phi$ with 
\begin{align} \label{eq:torque_bulk}
\frac{\tau_{\rm B}}{H^2} &= \tilde{\chi}_u(H) f_u(\theta) + \tilde{\chi}_s(H) f_s(\theta)+ \tilde{\chi}_\Phi(H) f_\Phi(\theta),
\end{align} 
The field-dependence of each contribution is described by the susceptibilities
 \begin{align}
 \tilde{\chi}_x(H) = \tilde{\chi}_{0,x} H^{-\zeta_x}
 \end{align}
  in terms of the constants $\tilde{\chi}_{0,x}$ and the scaling exponents $\zeta_{x}$, with $x=\{u,s,\Phi\}$. Since the effective uniform and staggered fields are orthogonal (see Supplementary Note 1), $\zeta_u$ and $\zeta_s$ are generally different. Constant susceptibility (i.e. $|\mathbf{m}| \propto H$) corresponds to the limit $\zeta \rightarrow 0$. The angle-dependence of the torque results entirely from the anisotropy of the effective fields, and is described by:
\begin{align}
f_x(\theta)&= -\frac{\text{d} \mathbf{h}_{\text{eff},x}}{\text{d} \theta} \cdot \frac{\mathbf{h}_{\text{eff},x}}{\vert \mathbf{h}_{\text{eff},x} \vert^{\zeta_x}}. \label{eq:f}
\end{align}
where $\mathbf{h}_{\text{eff},x}=\mathbf{H}_{\text{eff},x}/H$.
These expressions define the torque up to the constants $\tilde{\chi}_{0,x}$ and the scaling exponents $\zeta_x$.
\\

\textbf{Bulk torque response of $\kappa$-Cu.} 
To compare with $\kappa$-Cu, we computed $\tau$ for fields rotated in the $ac^*$-plane, and a variety of possible $\tilde{\chi}_{0,x}$ and $\zeta_x$ using  Eq.~\eqref{eq:torque_bulk} together with {\it ab-initio} values for the $g$-tensors and DM-interactions (see Supplementary Table~I). Coordinates were defined similarly to the experiment of Isono {\it et~al.}~\cite{isono2016quantum}, in which
$\theta$ is the angle between $\mathbf{H}$ and the long axis of an
ET molecule, approximated by the principal axis of $(\mathbb{G}_u \cdot
\mathbb{G}_u^{\rm T})$ close to the $a$ axis (see Fig.~\ref{fig:setup}a).

The experimental response\cite{isono2016quantum} for $T>T^*$ is consistent with a material deep in a vanilla QSL or paramagnetic state that corresponds to $\zeta_u=0$, $\zeta_s=0$ and $\tilde{\chi}_\Phi = 0$, i.e. the spin susceptibilities are field-independent and there is no chiral response (Fig.~\ref{fig:staggered}a). In this case, the uniform contribution $\tau_u$ dominates, and the torque arises primarily from weak anisotropy of the uniform $g$-tensor $\tilde{\mathbb{G}}_u$, providing a simple $\sin2 (\theta-\theta_0)$ dependence with a fixed $\theta_0 \approx 90^\circ$, as defined in Fig.~\ref{fig:staggered}a. 

For $T<T^*$, the experiments of Isono {\it et~al.} instead revealed a field dependent torque susceptibility that diverges at low temperatures as $\tau/H^2 \propto H^{-\zeta}\sin2(\theta-\theta_0)$, with $\zeta=0.76 -0.83$, together with a notable angle shift of $\theta_0 (H,T)$. Importantly, we find that these observations cannot be reconciled with divergence of any bulk susceptibility. For example, the case of a diverging {\it uniform} susceptibility $\tilde{\chi}_u(H)$ (Fig.~\ref{fig:staggered}b, $\zeta_u = 0.8$), then $\tau_u$ would dominate at all $T$, thus providing a nearly identical angle dependence at all $T$, with no shifting $\theta_0$. This scenario would also imply proximity to a ferromagnetic instability, and therefore appears unlikely given the strong antiferromagnetic interactions\cite{motrunich2005variational,qi2009dynamics,chubukov1994theory,chubukov1994quantum,sandvik2002striped,sandvik2011thermodynamics,kim1997massless}.

In this context, divergence of the {\it staggered} susceptibility $\tilde{\chi}_s(H)$ appears more likely, which may occur near a critical point
between a QSL and a N\'eel phase. We have previously
highlighted this scenario~\cite{winter2017SOC} as a possible explanation of the
$\mu$SR response of $\kappa$-Cu. However, we find this scenario is also inconsistent with the experimental torque. Figs.~\ref{fig:staggered}c,d depict the torque for an exponent $\zeta_s=0.8$, for different values of $\tilde{\chi}_s/\tilde{\chi}_u$. For a diverging staggered susceptibility, $\tau_s$ shows a saw-tooth like angle dependence instead of the experimentally observed $\sin2 (\theta-\theta_0)$. This behaviour stems from the strong angle dependence of the effective staggered field $\vert \mathbf{H}_{\text{eff},s} \vert$, 
which contributes to the denominator of Eq.~\eqref{eq:f}. The strongly anisotropic staggered $g$-tensor $\tilde{\mathbb{G}}_s$ leads to field orientations where $\vert \mathbf{H}_{\text{eff},s} \vert$ vanishes linearly as $\vert \mathbf{H}_{\text{eff},s} \vert \sim (\theta-\theta_c)$. For such orientations, a diverging susceptibility $ {\rm d}^2 E /{\rm d} \vert \mathbf{H}_{\text{eff},s} \vert^2$ implies divergence of ${\rm d}\tau/{\rm d}\theta \sim {\rm d}^2 E / {\rm d} \theta^2$, thus giving a saw-tooth torque.

Similar considerations apply to the case of a diverging {\it chiral}
susceptibility, shown in
Figs.~\ref{fig:staggered}e,f (with $\mathbf{n} || \mathbf{a}$ and $\zeta_\phi = 0.8$). In this case, the saw-tooth form reflects the vanishing of $\langle
\mathcal{H}_{\chi,\text{eff}} \rangle$ for in-plane fields $\mathbf{H}\perp
\mathbf{n}$, which leads to similar divergences in $\rm{d}\tau/\rm{d}\theta$
for $\zeta_\Phi > 0$. Taken together, the observed field- and temperature dependent angle shift $\theta_0$ and absence of saw-tooth features~\cite{isono2016quantum} argue against any diverging bulk susceptibility in $\kappa$-Cu. 
\\

\textbf{Local spin defects in $\kappa$-Cu.} 
In order to account for the experimental torque observations at $T<T^*$, we considered additional contributions from rare local spin moments induced by disorder, which may also give rise to diverging susceptibilities\cite{igloi2005strong,vojta2010quantum,lin2003low,vojta2006rare}. We identify two relevant types of disorder for $\kappa$-Cu. 

\begin{figure}
\includegraphics[width=\columnwidth]{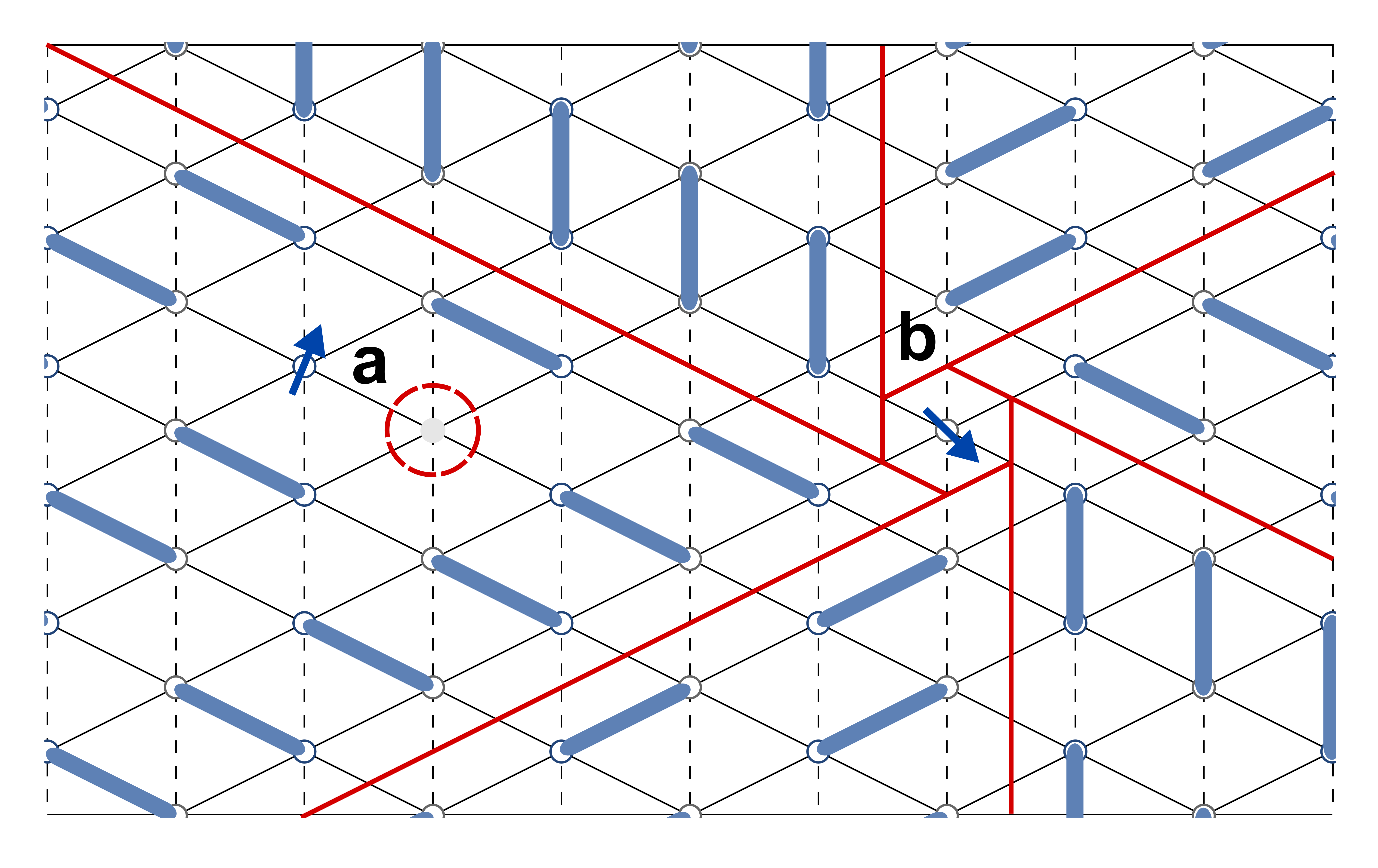}
\caption{\textbf{Local valence bond defects.} Domain walls between valence bond patterns are illustrated in red. (a) Local spin 1/2 caused by the breaking of a singlet bond due to an anion layer vacancy, emphasized by the red circle. (b) Local spin 1/2 due to a defect in the valence bond pattern.} \label{fig:imp_scheme}
\end{figure}

The first type produces a random modulation of the magnetic interactions between dimers, and includes disorder in the conformations of the terminal ethylene groups of the ET molecules\cite{guterding2015influence,hartmann2014mott}, the orientations of the cyanide anions located at crystallographic
inversion centers\cite{pinteric2014anisotropic}, and/or the local charge distributions within each dimer\cite{hotta2010quantum,abdel2010anomalous,kaneko2017charge}. In one-dimensional spin chains, introducing a finite randomness induces a ``random singlet'' ground state\cite{dasgupta1980c,fisher1994random}, in which the fluctuating singlets of the QSL are randomly pinned to form quasi-static singlets with varying length scales. At any given energy scale, a small fraction of spins remain unpaired, which leads to a diverging susceptibility. As recently emphasized by Kimchi {\it et
al.}~\cite{kimchi2018heat,kimchi2017valence}, similar behavior may occur in higher dimensional analogues of the random singlet state, which we refer to as valence bond glass (VBG)\cite{tarzia2008valence,singh2010valence} states. A possible scenario\cite{liu2018random} occurs in proximity to a valence bond solid order. In this case, random interactions induce complex patterns of domain walls that can host quasi-free orphan spins at their intersections, as depicted in Fig.~\ref{fig:imp_scheme}b. 

The second type of disorder may result from defects in the anion
layer\cite{drozdova2001composition,padmalekha2015esr}, which may slightly dope the
organic layer, producing non-magnetic spin vacancies. As illustrated in Fig.~\ref{fig:imp_scheme}a, these vacancies  break singlet bonds, which may produce local moments if the host system is in a confined state (e.g. a valence bond solid)\cite{vojta2000quantum,sachdev1999quantum,sachdev2003quantum,doretto2009quantum,kolezhuk2006theory,hoglund2007anomalous}.  Interestingly, Furukawa
{\it et~al.}\cite{furukawa2015quantum} recently showed that the introduction of anion defects via irradiation could suppress magnetic order in the less magnetically frustrated $\kappa$-(ET)$_2$Cu[N(CN)$_2$]Cl salt. 

Regardless of their origin, the low energy response of disorder-induced orphan spins has been the subject of many recent works\cite{liu2018random,westerberg1995random,westerberg1997low,lin2003low,igloi2005strong,poilblanc2010impurity}, including those with specific reference to $\kappa$-Cu\cite{watanabe2014quantum,shimokawa2015static}. In the following, we start from the assumption that such local moments exist at low energies, and are randomly distributed in the material.  We assume that the orphan spins interact via random long-range interactions that arise from bulk fluctuations, and consider their contribution to the magnetic torque.\\

\textbf{Defect torque expressions.}
For simplicity, we consider the case of $S=1/2$ impurity spins embedded in a non-ordered background, and assume constant bulk susceptibilities $\tilde{\chi}_u$ and $\tilde{\chi}_s$. Due to fluctuations of the valence bonds around each impurity, the associated magnetic moment will be delocalized over some characteristic localization length. An important distinction from the bulk case is that all Fourier components of the induced defect magnetization density are parallel or antiparallel at lowest order\cite{sachdev2003quantum}. Therefore, the impurity induced magnetism can be described by a single effective impurity spin variable $\tilde{\mathbf{S}}_{\rm I}$, which represents both the impurity and the broad screening cloud surrounding the impurity. An external field couples to $\tilde{\mathbf{S}}_{\rm I}$ through the uniform and staggered moments induced near the impurity. These are given by $\sum_{i^\prime\sim m}\langle \tilde{\mathbf{S}}_{i^\prime} \rangle 
= c_u \langle \tilde{\mathbf{S}}_{\text{I},m} \rangle$ and 
$\sum_{i^\prime\sim m} \eta_{i^\prime}\langle \tilde{\mathbf{S}}_{i^\prime} \rangle 
= c_s \langle \tilde{\mathbf{S}}_{\text{I},m} \rangle$ with $\eta_{i^\prime}=\pm1$, respectively. Here, the summation runs over dimer sites $i^\prime$ near the impurity $m$, and $c_u$ and $c_s$ are constants related to the Fourier transform of the induced spin density at $k=0$ and $k=(\pi,\pi)$, respectively. At lowest order, the scalar spin chirality is not coupled to the impurity effects. Therefore only the staggered and uniform moments respond to an external field through:
\begin{align}
\mathcal{H}_{\rm Zee, I} = -\mu_B \sum_m \mathbf{H}\cdot \tilde{\mathbb{G}}_{\rm I} \cdot  \tilde{\mathbf{S}}_{\text{I},m} 
\end{align}
where the effective impurity $g$-tensor is $\tilde{\mathbb{G}}_\text{I}\approx( c_u  \tilde{\mathbb{G}}_u + c_s \tilde{\mathbb{G}}_s )$, and therefore differs from any of the bulk $g$-tensors. The effective field felt by each impurity is given by:
\begin{align}
\mathbf{H}_{\text{eff},\rm I} = \tilde{\mathbb{G}}_{\rm I}^\text{T} \cdot \mathbf{H}
\end{align}
with the corresponding reduced field $\mathbf{h}_{\text{eff},\rm I} = \mathbf{H}_{\text{eff},\rm I}/H$. For simplicity, we take $\tilde{\mathbb{G}}_{\rm I}$ to be the same for each impurity.

In the absence of residual interactions between $S=1/2$ impurities, they would act as independent Curie spins, with the impurity contribution to the torque $\tau_{\rm I}/H^2$ diverging as $T^{-1}$ and $H^{-1}$ at low-field and temperature. However, the overlap of the screening clouds leads to random residual interactions $\{J_{mn}^{\rm eff}\}$ between the
randomly distributed impurity moments $m$ and $n$. Following previous works\cite{westerberg1995random,westerberg1997low,lin2003low,igloi2005strong}, we expect the interacting impurities to become successively coupled into clusters as the energy is lowered below the interaction energy (e.g. $k_BT \lesssim \text{max}|J_{mn}^{\rm eff}|$). The initial distribution of
interactions
may include both ferromagnetic ($J_{mn}^{\rm eff} < 0$) and antiferromagnetic ($J_{mn}^{\rm eff} > 0$) terms, as the sign of the coupling depends on the relative positions of the impurities. 
Provided this distribution
is not too singular\cite{westerberg1997low}, the low-energy response of the
impurities is therefore expected to be described by a finite randomness ``large-spin'' fixed point (LSFP)~\cite{westerberg1995random,westerberg1997low} or the ``spin-glass''
fixed point (SGFP)~\cite{lin2003low}. In the vicinity of such a fixed point, we find that the total impurity torque can be approximated by a modified Brillouin function $\mathcal{B}(S,x)$  (see Supplementary Note 3):
\begin{align} \label{eq:magI_gen}
\frac{\tau_{\rm I}}{H^2} \approx  g(\theta) \frac{N_C S_{\rm eff}^{\rm avg}}{H}
\mathcal{B}\left(S_{\rm eff}^{\rm avg},\frac{\mu_B| \mathbf{H}_{\text{eff},\rm I} |}{k_B T} \right),\\
g(\theta) = -\left(\frac{\text{d} \mathbf{h}_{\text{eff},\rm I}}{\text{d} \theta} \cdot\frac{ \mathbf{h}_{\text{eff},\rm I}}{|\mathbf{h}_{\text{eff},\rm I}|} \right)
\end{align}
where $N_C =N_0 \Omega^{2\kappa}$ is the number of clusters, and $S_{\rm eff}^{\rm avg} =S_0 \Omega^{-\kappa}$ is the average moment per cluster. Both depend on an effective energy scale defined by:
\begin{align} \label{eq:omega}
\Omega = \max ( k_B T, \mu_B S_{\text{eff}}^{\text{avg}} \vert \mathbf{H}_{\text{eff},\rm I} \vert).
\end{align}
The constants $S_0$ and $N_0$ define the average cluster spin and density at fixed $T$ and $H$. The non-universal exponent $\kappa$ is related to the fixed point distribution of energy couplings, and is sample-dependent but constrained by $1 \lesssim (2\kappa)^{-1} \leq \infty$ (see Supplementary Note 3). 

Evaluating the torque in the high-field limit $k_BT \ll \mu_B S_{\rm eff}^{\rm avg}| \mathbf{H}_{\text{eff},\rm I}|$, yields that $\tau_{\rm I}$ is temperature independent:
\begin{align}
\frac{\tau_I}{H^2} \approx & \ \tilde{\chi}_{\rm I}^H(H) f_{\rm I}^H(\theta)  \label{eq:torqueIH}\\
\tilde{\chi}_{\rm I}^H(H) = & \ \tilde{\chi}_{0, \rm I}^H H^{-\zeta_{\rm I}} \label{eq:torqueIH2}
\end{align}
where $\tilde{\chi}_{0,\rm I}^H$ is a constant that depends on the parameters $\{N_0,S_0,\zeta_{\rm I}\}$ (see Supplementary Note~3). 
The angle dependence is given by:
\begin{align}
f_{\rm I}^H(\theta)&= -\frac{\text{d} \mathbf{h}_{\text{eff},\rm I}}{\text{d} \theta} \cdot \frac{\mathbf{h}_{\text{eff},\rm I}}{\vert \mathbf{h}_{\text{eff},\rm I} \vert^{\zeta_{\rm I}}} 
\end{align}
The susceptibility diverges as $\tilde{\chi}_{\rm I}^H \propto H^{-\zeta_{\rm I}}$, with the exponent $\zeta_{\rm I} = (1+\kappa)^{-1}$ being restricted to the narrow range $2/3 \lesssim \zeta_{\rm I} \leq 1$ due to the constraints for the non-universal exponent $\kappa$. As a result, the low temperature divergence of $\tau_{\rm I}/H^2$ with respect to field is always marginally weaker than free Curie spins.

In the low-field limit $ \mu_B S_{\rm eff}^{\rm avg}| \mathbf{H}_{\text{eff},\rm I}| \ll k_BT$, the torque susceptibility evaluates to:
\begin{align} \label{eq:torqueIT}
\frac{\tau_{\rm I}}{H^2} \approx & \  \tilde{\chi}_{\rm I}^T(T) f_{\rm I}^T(\theta) \\
\label{eq:torqueIT2}
\tilde{\chi}_{\rm I}^T(T) = & \ \tilde{\chi}_{0,\rm I}^T \left(\frac{1}{k_BT}+S_0^{-1}(k_BT)^{\frac{1}{\zeta_{\rm I}}-2}\right)
\end{align}
which is independent of field. The derivation of the temperature-independent contribution $\tilde{\chi}_{0, \rm I}^T$ is given in Supplementary Note~3. The angle dependence $f_{\rm I}^T(\theta)$ is described by:
\begin{align}\label{eq:M4}
f_{\rm I}^T(\theta)&= -\frac{\text{d} \mathbf{h}_{\text{eff},\rm I}}{\text{d} \theta} \cdot \mathbf{h}_{\text{eff},\rm I} 
\end{align}
As discussed in the next section, despite not following a perfect power law, the low-field divergence of $\tau_{\rm I}/H^2$ with respect to temperature may appear to follow a power law $\sim T^{-\omega_{\rm I}}$ with $2-\zeta_{\rm I}^{-1}\leq \omega_{\rm I} \leq 1$ at intermediate temperatures.
\\

\begin{figure*}
\includegraphics[width=\textwidth]{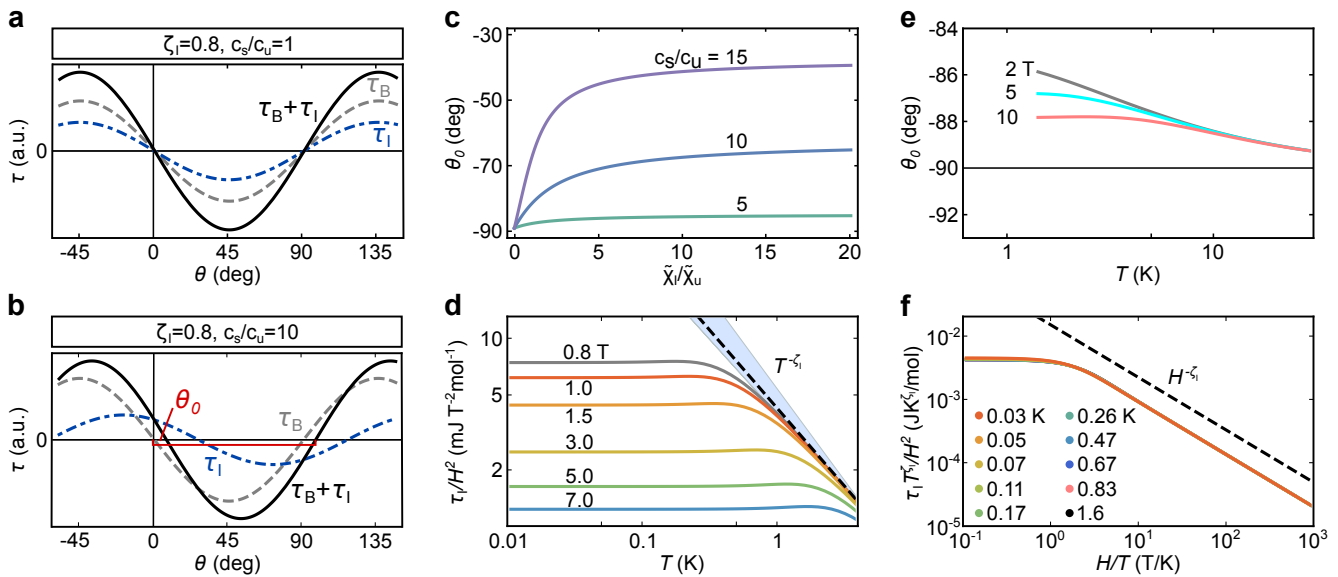}
\caption{\textbf{Physical quantities derived from magnetic torque considering effects of defect spins.} (a,b) Angle-dependence of the magnetic torque for $\tilde{\chi}_{\rm I}/\tilde{\chi}_u = 1$ and $\zeta_{\rm B}=0$ for different ratios of $c_s/c_u$. The total torque is indicated in solid black, the bulk contribution in dashed gray and the impurity contribution in dotted-dashed blue lines. (c) Angle shift in terms of $\tilde{\chi}_{\rm I}/\tilde{\chi}_u$, (d) $\tau_{\rm I}/H^2$ as a function of temperature for various magnetic fields (compare with Ref.~\onlinecite{isono2016quantum}, Fig.~4(d)), (e) Angle shift as a function of temperature for various magnetic fields (compare with Ref.~\onlinecite{isono2016quantum}, Fig.~2(c)), (f) plot of $\tau_{\rm I}T^{\zeta_{\rm I}}/H^2$ showing apparent $H/T$ scaling over several orders of magnitude (compare with Ref.~\onlinecite{isono2016quantum}, Fig.~5(a)). Plots (d)-(f) employ parameters from fitting the experimental data.} \label{fig:impurity_torque}
\end{figure*}


\textbf{Total torque response of $\kappa$-Cu.}
Here, we show that impurity-related orphan spin contributions can reproduce {\it all} observed features of the torque response. 

In Figs.~\ref{fig:impurity_torque}a,b we show the low temperature angle-dependence of the torque for $\mathbf{H}$ rotated in the $ac^\ast$-plane. Including impurity contributions, the total experimental torque is $
\tau=\tau_{\rm B}+\tau_{\rm I}$ according to 
Eq.~\eqref{eq:torque_bulk} and Eq.~\eqref{eq:torqueIH}, respectively.  We employed {\it ab-initio} values (given in Supplementary Table~I), and an impurity exponent $\zeta_{\rm I}=0.8$. We assume that the bulk susceptibility is not diverging (i.e. $\zeta_u = \zeta_s = \zeta_\Phi = 0$), and assume the same order of magnitude $\tilde{\chi}_{\rm
I,0}^H=\tilde{\chi}_{0,u}=\tilde{\chi}_{0,s}$ for simplicity. In agreement with the measurements for
$\kappa$-Cu, the total torque has a sinusoidal $\sin2(\theta-\theta_0)$ shape. Owing to a nearly isotropic impurity $g$-tensor 
$\tilde{\mathbb{G}}_{\rm I}$, a diverging $\tau_{\rm I}/H^2$ does not
lead to a saw-tooth appearance. However, provided the existence of a finite Dzyaloshinskii-Moriya interaction, then $\tilde{\mathbb{G}}_{\rm I}$ will generally differ from the bulk $\tilde{\mathbb{G}}_u$, leading to a shifted angle dependence of the impurity contribution. This explains the experimentally observed\cite{isono2016quantum} field and temperature dependent angle shift $\theta_0(H,T)$. As illustrated in Fig.~\ref{fig:impurity_torque}c, the value of $\theta_0$ measures the ratio of impurity and bulk susceptibilities $\tilde{\chi}_I/\tilde{\chi}_u $. The asymptotic value is further
controlled by the ratio of the constants $c_s/c_u$, corresponding to the staggered ($\mathbf{k}=(\pi,\pi)$) and uniform ($\mathbf{k}=0$) contributions of the induced spin density. For $c_s/c_u=1$ (Fig.~\ref{fig:impurity_torque}a) the impurity and bulk
contributions have nearly identical angle dependence, since
$\tilde{\mathbb{G}}_{\rm I} \approx \tilde{\mathbb{G}}_{u}$. This leads to $\theta_0 \sim -90^\circ$. In
contrast, for $c_s/c_u>1$ (Fig.~\ref{fig:impurity_torque}b), the impurity contribution is shifted with respect to
the bulk contribution (Fig.~\ref{fig:impurity_torque}b), which leads to an overall shift of the total torque curve.  The experimentally observed angle shift\cite{isono2016quantum} corresponds to $c_s/c_u \approx 5-10$, implying the staggered moment induced around each defect exceeds the uniform defect moment.

To produce Figs.~\ref{fig:impurity_torque}d-f, we performed a global fit of the divergent experimental\cite{isono2016quantum} torque at various $T,H$, yielding values for $N_0, S_0$ and $\zeta_{\rm I}$ in Eq.~\eqref{eq:magI_gen}.
 The fitted exponent ($\zeta_{\rm I} = 0.79 \pm 0.03$) falls in the middle of the suggested range $2/3 \lesssim \zeta_{\rm I} \leq 1$, and is therefore consistent with a scenario with orphan spins. Fig.~\ref{fig:impurity_torque}d shows the resulting temperature dependence at various fields. Consistent with the experiment, the theoretical $\tau_{\rm I}/H^2 \sim T^{-\omega_{\rm I}}$ appears to follow a power law behaviour.
From Eq.~\eqref{eq:torqueIT2} it follows that the exponent  
should fall in the narrow range $2-\zeta_{\rm I}^{-1}\leq \omega_{\rm I} \leq 1$, illustrated by the blue region in Fig.~\ref{fig:impurity_torque}d. For this reason, $\omega_{\rm I}\approx \zeta_{\rm I}$ (dashed line) at low-field. Interestingly, the similar values of $\zeta_{\rm I}$ and $\omega_{\rm I}$ mean that $\tau_{\rm I}$ will accidentally {\it appear} to display $H/T$ scaling described by $\tau_{\rm I}/H^2 \approx T^{-\zeta_{\rm I}}F[H/T]$, with:
\begin{align}
F[X] = \left\{\begin{array}{cc}\text{constant} & X \ll 1 \\ X^{-\zeta_{\rm I}} & X \gg 1 \end{array} \right.
\end{align}
This behaviour is displayed in Fig.~\ref{fig:impurity_torque}f, where the theoretical $(\tau_{\rm I}/H^2)T^{\zeta_{\rm I}}$ is plotted against $(H/T)$ for different temperatures. Deviations from scaling appear as a very small separation of the curves at low-fields. Finally, in Fig.~\ref{fig:impurity_torque}e we show the predicted evolution of the angle shift $\theta_0$, for a fixed ratio of induced staggered to uniform defect moments $c_s/c_u = 6$. At high temperature, $\theta_0$ asymptotically approaches $90^\circ$, as thermal fluctuations suppress the diverging impurity contribution.

Taken together, the impurity contributions appear to explain all essential features of the experimental torque response below $T^\ast$, including the reported range of exponents $\zeta_{\rm exp} = 0.76 - 0.83$, the $\sin 2(\theta-\theta_0)$ dependence, the evolution of the angle shift $\theta_0$, and the apparent $H/T$ scaling of the magnitude of the torque. 
\\

\textbf{Inhomogeneous NMR response.}
Within the same framework, we have also considered the inhomogeneous broadening of the NMR lines in $\kappa$-Cu, which has previously been attributed to impurity spins\cite{shimizu2006emergence,gregor2009nonmagnetic}. As the external field aligns the total spin of each cluster, the local defect moments $\tilde{\mathbf{S}}_{\text{I},m}$ contributing to the cluster become static on the NMR timescale. The resulting static spin density around each defect is staggered, and decays with distance, leading to an inhomogeneous distribution of staggered Knight shifts within the sample. At first approximation, we assume that the resulting contribution to the NMR linewidth $\nu_{\rm I}$ scales as the root mean squared average of $\tilde{\mathbf{S}}_{\text{I},m}$, leading to (see Supplementary Note 3):
\begin{align} \label{eq:NMRline}
\nu_{\rm I} \approx \nu_0 \  \mathcal{B}\left(S_{\rm eff}^{\rm avg},\frac{\mu_B|\mathbf{H}_{\text{eff},\rm I}|}{k_B T} \right).
\end{align}
For the low-temperature or high-field limit $k_BT \ll \mu_B S_{\rm eff}^{\rm avg}| \mathbf{H}_{\text{eff},\rm I}|$, the local impurities are completely static, leading to constant linewidth ($\nu_{\rm I} = \nu_0$). In the opposite limit $k_BT \gg \mu_B S_{\rm eff}^{\rm avg}| \mathbf{H}_{\text{eff},\rm I}|$, expanding the Brillouin function yields:
\begin{align}\label{eq:NMR_imp}
\nu_{\rm I} \propto\frac{H}{T^{1/\zeta_{\rm I}} }\left(1+S_0^{-1}(k_BT)^{\frac{1-\zeta_{\rm I}}{\zeta_{\rm I}}}\right).
\end{align}
suggesting $\nu_{\rm I}$ is linear with respect to field. 

\begin{figure}
\includegraphics[width=\columnwidth]{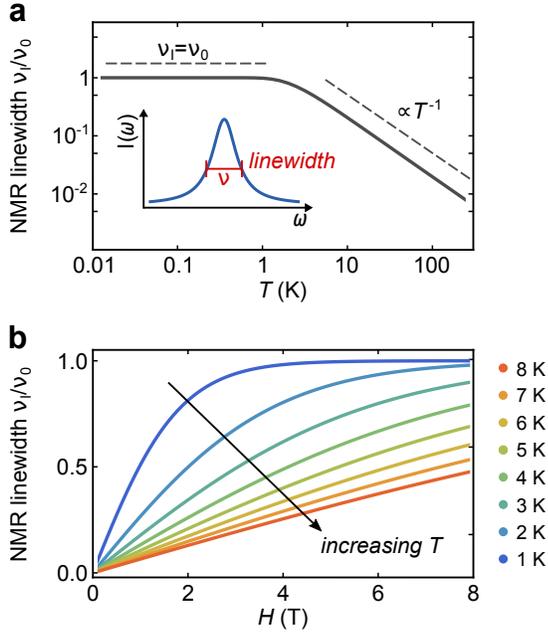}
\caption{\textbf{Relative impurity contribution to the $^{13}$C NMR linewidth}. $\nu_{\rm I}/\nu_0$ according to Eq. \eqref{eq:NMRline} as a function of (a) temperature for fixed field $H = 4$ T, and (b) as a function of $H$ for various temperatures.} \label{fig:imp_NMR}
\end{figure}

In Fig.~\ref{fig:imp_NMR} we show the predicted relative linewidth $\nu_{\rm I}/\nu_0$ as a function of temperature and field employing $S_0,N_0$ and $\zeta_{\rm I}$ fit from the experimental $\tau$. Comparing Eq.~\eqref{eq:NMR_imp} to the experimental behaviour, we find reasonable agreement. The experimental linewidth saturates  (Ref.~\onlinecite{shimizu2006emergence}, Fig.~2) in precisely the same temperature range where $\tau/H^2$ also becomes temperature independent (Ref.~\onlinecite{isono2016quantum}, Fig.~3). For $T>1$ K, the experimental NMR linewidth also grows approximately linear with $H$  (Ref.~\onlinecite{shimizu2006emergence}, Fig.~4). This correspondence suggests that the NMR linewidth and torque response have a common impurity origin.

%

\section{Discussion} 

In the presence of anisotropic interactions beyond the conventional Heisenberg terms, measurements of magnetic torque $\tau$ may provide significant insights into quantum magnets. We have shown that the effects of spin-orbit coupling
allow the torque to probe uniform and staggered components of spin susceptibilities. In addition, higher order ring-exchange
processes, resulting from proximity to a metallic state, directly couple the magnetic field to the scalar spin chirality. For this reason,
the torque provides a direct probe for a number of possible instabilities of the spin-liquid state in $\kappa$-Cu, including
proximity to magnetically ordered $(\pi,\pi)$ N\'eel or chiral ordered phases.
However, we showed that the experimental response at low temperature is incompatible with the divergence of any bulk susceptibility, arguing against proximity to any such instability.

We therefore considered contributions to $\tau/H^2$ from rare 
impurity spins. After deriving
approximate expressions for such contributions, we find that effects of local ``orphan'' spins can explain essentially {\it all}
features of the torque experiments in $\kappa$-Cu, including the specific angle dependence, observed exponents, and the apparent $H/T$ scaling noted
by Isono {\it et~al.}\cite{isono2016quantum}. Related expressions also consistently describe the inhomogeneous NMR linewidth~\cite{shimizu2006emergence} with the same parameters. These independent experiments both point to the presence of local moments in $\kappa$-Cu, which we attribute to a disorder-induced valence bond glass (VBG)\cite{tarzia2008valence,singh2010valence} ground state analogous to the random-singlet phase of 1D spin chains\cite{dasgupta1980c,fisher1994random}. While the small fraction of orphan spins result in a diverging torque response, the majority of low-energy
excitations will consist of local domain wall fluctuations, i.e.~shifts of the
frozen valence bond pattern\cite{singh2010valence}. Such fluctuations can give rise to a linear
specific heat \cite{frischmuth1999thermodynamics,watanabe2014quantum} $C_v \sim T$, in analogy with structural
glasses\cite{doi:10.1080/14786437208229210}. Since most excitations of
the valence bond glass are ultimately localized, they will contribute
negligibly to thermal conductivity $\kappa_T$ in the $T \to 0$ limit. This feature
may therefore explain the linear $C_v$ but suppressed $\kappa_T$ observed in
$\kappa$-Cu\cite{yamashita2009thermal,yamashita2012thermal}. Since disorder is expected to be small compared to the scale of interactions, the appearance of local moments also places some constraints on the ground state of the hypothetical disorder-free sample\cite{vojta2000quantum,sachdev1999quantum,sachdev2003quantum,doretto2009quantum,kolezhuk2006theory,hoglund2007anomalous}. In particular, proximity to valence bond solid order may aid the formation of a VBG\cite{liu2018random}. 

We note, however, that an important discrepancy between the scaling expressions and the experimental response occurs at the $T^* =$ 6 K anomaly. At $T^*$, both the experimental torque angle shift $\theta_0$ and the NMR linewidth $\nu$ increase more sharply than suggested by the scaling ans\"atze, so that strong contributions from orphan spins seem to emerge rapidly at $T^*$. We therefore speculate that the anomalies observed in a wide range of
experiments are connected to the freezing
of valence bonds into the random (VBG) configuration. 
In principle, this may be driven by a number of effects. 
For example, it may reflect a thermal VBS ordering transition in hypothetical disorder free samples, which evolves into a VBG transition with finite disorder. Alternately, external factors such as freezing of local charges or anion orientations may lead to a rapid enhancement of the effective disorder, driving the rapid formation of the VBG. In either case, the singlets would be free to fluctuate for $T\gg T^*$, thus forming an essentially homogeneous QSL at high temperatures. Strong
inhomogeneity in the NMR relaxation (and $\mu$SR response\cite{nakajima2012microscopic}) would only be expected below $T^*$, which is consistent with
Ref.~\onlinecite{shimizu2006emergence}. Moreover, the temperature $T^*$ itself should not be strongly sensitive to
external fields, which is consistent with negligible field
dependence\cite{manna2010lattice} of the thermal expansion near $T^*$.

On this basis, we conclude that the torque investigations of $\kappa$-Cu provide unique insight into the low temperature phase. The analysis presented in this work provide direct evidence of disorder-related effects and may provide a roadmap for finally revealing the nature of the enigmatic $T^*$ anomaly as the formation of a valence bond glass. This framework may also be generalized to other prominent organic QSL candidates like $\kappa$-H$_3$(Cat-EDT-TTF)$_2$\cite{isono2014gapless} or $\beta^\prime$-EtMe$_3$Sb[Pd(dmit)$_2$]$_2$\cite{watanabe2012novel}. It is also noteworthy that similar scaling of the susceptibility with exponent $\zeta = 2/3$ is also observed in the frustrated Kagome system Herbertsmithite\cite{helton2010dynamic}, where disorder is thought to play a role. This highlights the rich interplay between disorder and frustration in quantum spin materials.


\begin{acknowledgments}
The authors acknowledge fruitful discussions with 
S. Brown, V. Dobrosavljevic, S. Hartmann, K. Kanoda,
I. Kimchi, M. Lang
and P. Szirmai.
The work was supported by the Deutsche Forschungsgemeinschaft (DFG) through project SFB/TRR49.
\end{acknowledgments}

\section*{Author contributions}
All authors made equal contributions.

\section*{Additional information}
\textbf{Competing interests:} The authors declare no competing financial interests.

\bibliographystyle{naturemag_noURL}
\bibliography{scaling}

\clearpage


\section*{Supplementary material}

# Supplemental Information for:
# Critical Spin Liquid versus Valence Bond Glass in Triangular Lattice Organic $\kappa$-(ET)$_2$Cu$_2$(CN)$_3$

Kira Riedl,[1] Roser Valentí,[1] and Stephen M. Winter*[1]

[1]*Institut für Theoretische Physik, Goethe-Universität Frankfurt,*
*Max-von-Laue-Strasse 1, 60438 Frankfurt am Main, Germany*

**SUPPLEMENTARY NOTE 1: SPIN HAMILTONIAN**

In this supplementary note, we present *ab-initio* calculations of the spin Hamiltonian for $\kappa$-Cu. In general, the magnetic interactions can be divided according to the number of spin operators appearing in each term $\mathcal{H}_{(n)} \sim \mathcal{O}(\mathbf{S}^n)$:

$$\mathcal{H} = \mathcal{H}_{(1)} + \mathcal{H}_{(2)} + \mathcal{H}_{(3)} + \mathcal{H}_{(4)} + ... \tag{S1}$$

with $\mathcal{H}_{(1)}$ including the Zeeman operators, $\mathcal{H}_{(2)}$ including bilinear spin interactions, etc. We first present calculations of the interactions in the crystallographic coordinate system ($\mathcal{H}$) up to fourth order, and then describe the effects of the local coordinate transformation $\mathcal{H} \to \tilde{\mathcal{H}}$ introduced in the main text, which removes the anisotropic spin interactions at lowest order. We have previously[1] estimated the nearest neighbour bilinear couplings using a combination of hopping integrals obtained from ORCA at the PBE0/def2-VDZ level and exact diagonalization of small clusters of molecules. Using this approach, we have extended the calculations to estimate also longer-range couplings and higher order ring-exchange terms.

The Zeeman operator can be generally written:

$$\mathcal{H}_{(1)} = -\sum_i \mathbf{H} \cdot \mathbb{G}_i \cdot \mathbf{S}_i \tag{S2}$$

in terms of the external field $\mathbf{H}$, local $g$-tensor $\mathbb{G}_i$ and spin operator $\mathbf{S}_i$ at dimer site $i$. Within each $P2_1/c$ unit cell, there are two molecular dimers, which are related by $2_1$ screw axes and $c$-glide planes, shown in Supplementary Figure 1(a). As mentioned in the main text, the $g$-tensors may differ for the two dimer sublattices, labelled A and B, by symmetry. For this reason, it is useful to divide the $g$-tensor into uniform $\mathbb{G}_u$ and staggered $\mathbb{G}_s$ components, with:

$$\mathbb{G}_i = \mathbb{G}_u + \eta_i \mathbb{G}_s \tag{S3}$$

$$\eta_i = \begin{cases} +1 & i \in \text{sublattice A} \\ -1 & i \in \text{sublattice B} \end{cases} \tag{S4}$$

In order to estimate $\mathbb{G}_u$ and $\mathbb{G}_s$ for $\kappa$-Cu, we performed density functional theory calculations on isolated dimers at the PBE0/IGLO-III level using the ORCA package[2,3]. The molecular geometry was taken from the 5 K crystal structure reported in Ref. 4. The principle axes $(p, q, r)$ of the local $g$-tensors for the A and B sublattices are illustrated in Supplementary

1

| $(g_p, g_q, g_r)$ | $J$ | $J'$ | $J''$ | $J'''$ | $(D_a, D_b, D_{c^*})$ | $K_h$ | $K_v$ | $K_d$ | $K'_h$ | $K'_v$ | $K'_d$ | $J_{\chi,(1T)}$ |
|---|---|---|---|---|---|---|---|---|---|---|---|---|
| $(2.002, 2.008, 2.010)$ | 228 | 268 | 9.5 | 5.1 | $(3.30, 0.94, 0.99)$ | 16.5 | 13.6 | -21.3 | 17.0 | 17.7 | -20.5 | -0.04 |

SUPPLEMENTARY TABLE 1. **Computed Hamiltonian Parameters.** $g$-tensor contribution along the principal axes $(p, q, r)$ and computed magnetic exchange interactions in K with respect to $(a, b, c^*)$, illustrated in Figs. 1 and 2.

Figure 1(b). The largest value $g_r = 2.010$ corresponds to the long axis of the molecules, while the second largest $g_q = 2.008$ lies along the axis connecting the two molecules within each dimer. The final value of $g_p = 2.002$ was found for the third principal axis. In the crystallographic $(a, b, c^*)$ coordinates, the uniform and staggered tensors are estimated as:

$$\mathbb{G}_u = \begin{pmatrix} 2.010 & 0 & -8 \cdot 10^{-4} \\ 0 & 2.005 & 0 \\ 7 \cdot 10^{-4} & 0 & 2.005 \end{pmatrix}, \tag{S5}$$

and:

$$\mathbb{G}_s = \begin{pmatrix} 0 & -4 \cdot 10^{-4} & 0 \\ -6 \cdot 10^{-4} & 0 & -26 \cdot 10^{-4} \\ 0 & -25 \cdot 10^{-4} & 0 \end{pmatrix}. \tag{S6}$$

At second order in the spin operators, the bilinear interactions can be generally written:

$$\mathcal{H}_{(2)} = \sum_{ij} J_{ij}\, \mathbf{S}_i \cdot \mathbf{S}_j + \mathbf{D}_{ij} \cdot (\mathbf{S}_i \times \mathbf{S}_j) + \mathbf{S}_i \cdot \Gamma_{ij} \cdot \mathbf{S}_j \tag{S7}$$

where $J_{ij}$ describes the Heisenberg coupling, $\mathbf{D}_{ij}$ is the Dzyaloshinskii-Moriya vector, and $\Gamma_{ij}$ is a traceless symmetric tensor describing the pseudo-dipolar interaction. We label the unique interactions according to Supplementary Figure 2(a). For example, the anisotropic triangular lattice of nearest neighbour bonds is composed of $J$ and $J'$ interactions, while longer range second neighbour couplings are labelled $J''$ and $J'''$.

Within the $P2_1/c$ space group, the presence of a crystallographic inversion center forbids a DM interaction between sites $i, j$ belonging to the same sublattice (i.e. $\mathbf{D}'_{ij} = 0$). Between different sublattices, $\mathbf{D}_{ij}$ is finite. By symmetry, the signs of the $a$ and $c^*$ axes components $D_a$ and $D_{c^*}$ are staggered with $k = (\pi, \pi)$ periodicity with respect to the square lattice bonds, as

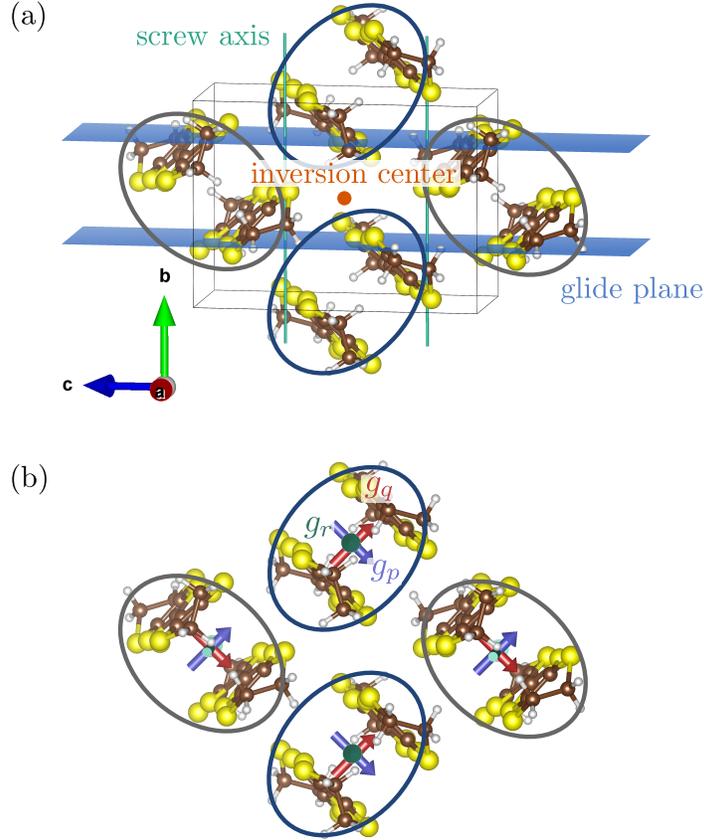

SUPPLEMENTARY FIGURE 1. **Symmetries in the $P2_1/c$ space group.** (a) Screw axes, glide planes and inversion center in the $P2_1/c$ space group, with two ET dimers per unit cell. (b) The principal axes $(g_p, g_q, g_r)$ are shown in the two sublattices, labelled A and B.

shown in Supplementary Figure 2(b). These components are responsible for the emergence of a small canted ferromagnetic moment in the magnetically ordered $(\pi, \pi)$-Néel phase of the related salt $\kappa$-(ET)$_2$Cu(N(CN)$_2$)Cl. In contrast, the $b$-axis component $D_b$ has a striped periodicity. This component does not couple to any of the magnetic states expected to be relevant for $\kappa$-phase salts; we have therefore ignored this component in first approximation.

At this point, it is convenient to introduce a site-dependent coordinate transformation discussed in the main text, and by e.g. Shekhtman *et al.*[5] Employing this transformation, it is possible to completely "gauge" away some components of the anisotropic interactions satisfying particular symmetries. In particular, if the DM vectors sum to zero around all closed loops on the lattice, it is possible to make site-dependent transformations of the spin coordinates that simultaneously eliminate *all* leading order anisotropic contributions to

3

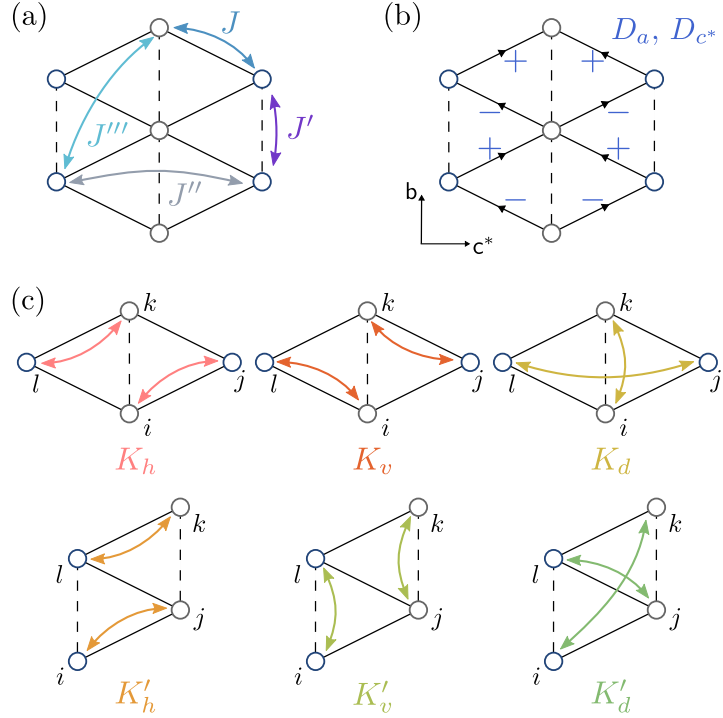

SUPPLEMENTARY FIGURE 2. **Definition of magnetic exchange parameters on anisotropic triangular lattice.** Each lattice point represents an ET dimer. (a) Heisenberg exchange parameters. (b) Pattern of $a$ and $c^*$ components of the DM interaction. $D_b$ follows a stripy pattern. (c) Definition of ring exchange parameters on the two distinct four site plaquettes in the anisotropic triangular lattice. As an example the spin Hamiltonian contains a term $K_v(\mathbf{S}_i \cdot \mathbf{S}_l)(\mathbf{S}_j \cdot \mathbf{S}_k)$ on a plaquette with one dashed and four solid bonds.

$\{\mathcal{H}_{(n)}\}$ for $n > 1$.

The proof of this possibility is rooted in the microscopic form of the hopping Hamiltonian of the underlying electronic system. It is convenient to express the hopping Hamiltonian in terms of the spinor operators $\mathbf{c}^\dagger = (\,c_\uparrow^\dagger\ c_\downarrow^\dagger\,)$ as:

$$\mathcal{H}_{\text{hop}} = \sum_{ij} e^{ia_{ij}} \mathbf{c}_i^\dagger \mathbb{T}_{ij} \mathbf{c}_j \tag{S8}$$

$$\mathbb{T}_{ij} = \left( t_{ij} \mathbb{I}_{2\times 2} + \frac{i}{2} \vec{\lambda}_{ij} \cdot \vec{\sigma} \right) \tag{S9}$$

Here, $\mathbb{I}_{2\times 2}$ is the $2 \times 2$ identity matrix, $t_{ij}$ is the spin-diagonal hopping, while off-diagonal hopping $\vec{\lambda}_{ij}$ arises as a result of spin-orbit coupling. The above restriction on the sum

of DM-vectors around any closed loop is equivalent to restricting the product of hopping matrices around any closed path to be a multiple of the identity matrix:

$$\mathbb{T}_{ij}\mathbb{T}_{jk}...\mathbb{T}_{lm}\mathbb{T}_{mi} = C\ \mathbb{I}_{2\times 2} \tag{S10}$$

We then consider making site-dependent spin rotations, which transform the operators as:

$$\tilde{\mathbf{c}}_i = e^{i\vec{v}_i\cdot\vec{\sigma}}\mathbf{c}_i \tag{S11}$$

in terms of some arbitrary vector $\vec{v}_i$. The transformed hopping matrices are then:

$$\tilde{\mathbb{T}}_{ij} = e^{-i\vec{v}_i\cdot\vec{\sigma}}\mathbb{T}_{ij}e^{i\vec{v}_j\cdot\vec{\sigma}} \tag{S12}$$

The question of interest is whether we can define a transformation, defined by a specific set of $\{\vec{v}_i\}$, such that $\tilde{\mathbb{T}}_{ij} = \tilde{t}_{ij}\mathbb{I}_{2\times 2}$ on every bond. In fact, the restriction of Eq. (S10) guarantees this possibility. To see this, consider starting at site $i$, and making a string of site-dependent transformations at sites $j, k, ...$ to bring $\tilde{\mathbb{T}}_{ij}, \tilde{\mathbb{T}}_{jk}, ...$ into diagonal form:

$$\tilde{\mathbb{T}}_{ij} = \mathbb{T}_{ij}e^{i\vec{v}_j\cdot\vec{\sigma}} = \tilde{t}_{ij}\ \mathbb{I}_{2\times 2} \tag{S13}$$

$$\tilde{\mathbb{T}}_{jk} = e^{-i\vec{v}_j\cdot\vec{\sigma}}\mathbb{T}_{jk}e^{i\vec{v}_k\cdot\vec{\sigma}} = \tilde{t}_{jk}\ \mathbb{I}_{2\times 2} \tag{S14}$$

This process can be repeated indefinitely until the loop is about to be closed. The global transformation is consistent only if the string of transformations is compatible with the last $\tilde{\mathbb{T}}_{mi}$ also being proportional to the identity. Since the product is invariant:

$$\mathbb{T}_{ij}\mathbb{T}_{jk}...\mathbb{T}_{lm}\mathbb{T}_{mi} = \tilde{\mathbb{T}}_{ij}\tilde{\mathbb{T}}_{jk}...\tilde{\mathbb{T}}_{lm}\tilde{\mathbb{T}}_{mi}, \tag{S15}$$

it holds that:

$$\tilde{\mathbb{T}}_{mi} = C\left(\prod_{i\to m}\frac{1}{\tilde{t}_{ij}}\right)\mathbb{I}_{2\times 2}. \tag{S16}$$

Therefore, the final hopping matrix is automatically made diagonal by this string of transformations, provided Eq. (S10) holds. This completes the proof for the existence of a transformation that sets all $\tilde{\lambda}_{ij} \to 0$. The practical implication is that SOC effects can be completely "gauged away" already at the level of the underlying hopping Hamiltonian, from which the spin couplings are derived. As a result, all interactions appearing in the transformed spin Hamiltonians $\{\tilde{\mathcal{H}}_{(n)}\}$ for $n > 1$ must take an isotropic form.

5

In the case of a staggered $(\pi, \pi)$ pattern, (as for the components $D_a$ and $D_{c^*}$ in $\kappa$-Cu), the local transformations $\mathbf{S} \to \tilde{\mathbf{S}}$ that eliminate the anisotropic couplings consist of rotations around $\mathbf{D}$ by the canting angle $\phi_i = 1/2\, \eta_i \arctan(|\mathbf{D}_{ij}|/J_{ij})$. Here, $\eta_i$ is defined as in Eq. (S4). This rotation is illustrated in the left panel with $H = 0$ of Supplementary Figure 3. To leading orders, the pseudo-dipolar tensor can be expressed in terms of the DM vector as $\Gamma \propto \mathbf{D} \otimes \mathbf{D}$. In this limit, $\Gamma$ is also exactly cancelled, leaving only the isotropic Heisenberg term:

$$\mathcal{H}_{(2),\text{eff}} = \sum_{ij} \tilde{J}_{ij}\, \tilde{\mathbf{S}}_i \cdot \tilde{\mathbf{S}}_j \tag{S17}$$

For small canting angles, i.e. weak spin-orbit coupling, we can work with the approximations $\cos\phi_i \approx 1$ and $\sin\phi_i \approx |\mathbf{D}_{ij}|/2J_{ij}$. This leads to $\tilde{J}_{ij} \approx J_{ij}$.

The utility of such a transformation is that it transfers the explicitly anisotropic terms to the Zeeman Hamiltonian, which contains then a uniform and staggered contribution:

$$\mathcal{H}_{\text{Zee,eff}} = -\mu_B \sum_i (\mathbf{H}_u + \eta_i \mathbf{H}_s) \cdot \tilde{\mathbf{S}}_i. \tag{S18}$$

For small canting angles, the two field terms become:

$$\mathbf{H}_u = \mathbb{G}_u^{\text{T}} \cdot \mathbf{H} \qquad \text{and} \qquad \mathbf{H}_s = (\mathbb{G}_s + \mathbb{R})^{\text{T}} \cdot \mathbf{H}, \tag{S19}$$

where we introduced the matrix:

$$\mathbb{R} = \frac{1}{2J} \mathbb{G}_u \cdot \begin{pmatrix} 0 & D_{c^*} & 0 \\ -D_{c^*} & 0 & D_a \\ 0 & -D_a & 0 \end{pmatrix} \tag{S20}$$

In the main text, these total terms are discussed as the total $g$-tensors in the rotated framework, which are for small canting angles:

$$\tilde{\mathbb{G}}_u = \mathbb{G}_u \qquad \text{and} \qquad \tilde{\mathbb{G}}_s = \mathbb{G}_s + \mathbb{R}. \tag{S21}$$

Note that due to the structure of $\mathbb{R}$ and of the $g$-tensors (S5) and (S6) the effective uniform and staggered field are orthogonal. This is relevant for the scaling behaviour of the field-induced uniform and staggered magnetization, introduced in the main text.

Similar transformations can be applied to study the 3-spin interactions $\tilde{\mathcal{H}}_{(3)}$. Since any product of three spins at different sites is odd under time-reversal, such interactions are forbidden at zero field. However, as mentioned in the main text, a finite magnetic flux through

6

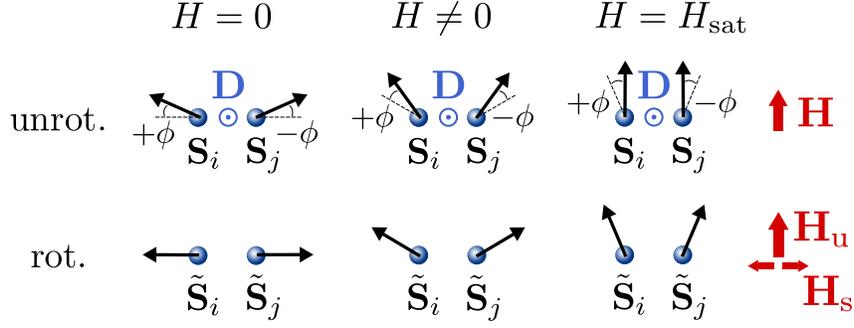

SUPPLEMENTARY FIGURE 3. **Gauging away anisotropic exchange terms.** Illustrated are the cases in the unrotated (upper panel) and rotated (lower panel) framework of the spins for zero, finite and saturation field. The effective staggered field $\mathbf{H}_s$ in the rotated framework is orthogonal to the uniform field $\mathbf{H}_u$.

the 3-site plaquettes can give rise to finite contributions to $\tilde{\mathcal{H}}_{(3)}$ that scales with odd powers of $|\mathbf{H}|$. These effects can be treated through the minimal coupling of the moving electron to the so-called Peierls phase $a_{ij} = \frac{q}{\hbar}\int_i^j \mathbf{A}\cdot d\mathbf{l}$ in the transformed hopping Hamiltonian

$$\tilde{\mathbb{T}}_{ij} \to \mathbf{c}_i^\dagger e^{ia_{ij}}\tilde{\mathbb{T}}_{ij} \qquad (S22)$$

This gives rise to odd order contributions in perturbation theory with a dependence on $\Phi = \oint_{\partial \mathcal{S}} \mathbf{A}\cdot d\mathbf{l}$. This quantity is independent of the local spin coordinates, and therefore is invariant under the transformation described above.

The dominant three-spin term is the so-called scalar spin chirality term,

$$\mathcal{H}_{(3),\text{eff}} = \frac{1}{S}\sum_{\langle ijk\rangle} \tilde{J}_\chi^{ijk}\, \tilde{\mathbf{S}}_i\cdot(\tilde{\mathbf{S}}_j\times\tilde{\mathbf{S}}_k). \qquad (S23)$$

with the exchange term given by (up to order $t^3$):

$$J_\chi^{(3)} = 24\frac{t_{ij}t_{jk}t_{ki}}{U^2}\sin\Phi, \qquad (S24)$$

where $\Phi$ is proportional to the magnetic flux enclosed by the triangular plaquette $\langle ijk\rangle$. With the assumption of a homogeneous magnetic field ($\mathbf{A} = \frac{1}{2}\mathbf{r}\times\mathbf{H}$) we may use the approximation

$$\Phi = \frac{q}{\hbar}\mu_B A_{\text{triangle}}\mathbf{H}^{\mathrm{T}}\cdot\mathbf{n}, \qquad (S25)$$

7

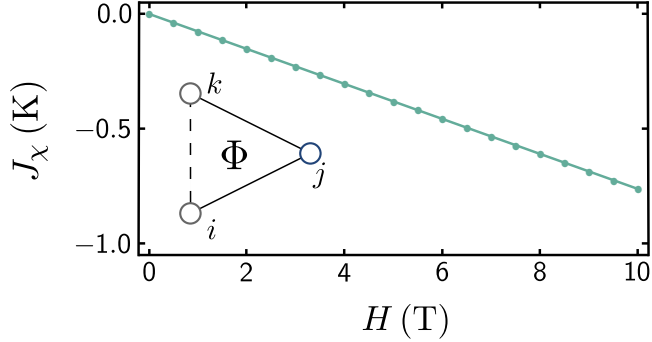

SUPPLEMENTARY FIGURE 4. **Exchange term of scalar spin chirality as a function of field.** The linear dependence on $H \propto \Phi$ is therefore confirmed numerically.

where $\mathbf{n}$ is the out-of-plane unit vector and $A_\text{triangle}$ the area formed by the triangular plaquette. Numerical estimates for $J_\chi^{ijk}$ as a function of field, are shown in Supplementary Figure 4, and given in Supplementary Table 1 for $H = 1\,\text{T}$. Following our previous approach, we performed exact diagonalization of the extended Hubbard Hamiltonian and projection on the corresponding low energy subspace on clusters of up to eight molecules. The hopping parameters used in the Hubbard picture were calculated with the ORCA package[2] and the two-particle parameters were chosen the same as in Ref. 1 with a Hubbard repulsion $U = 0.55\,\text{eV}$, a Hund's coupling $J_\text{H} = 0.2\,\text{eV}$, and a nearest neighbour Hubbard repulsion $V = 0.15\,\text{eV}$. Here, we considered the largest term, corresponding to triangles with two $J$-bonds, and one $J'$-bond. In the limit of small fluxes $\sin\Phi \approx \Phi$, so that the exchange term depends for $\mathcal{O}(t^3)$ linearly on the field.

For convenience, it is useful to write the Hamiltonian in analogy with a Zeeman term:

$$\mathcal{H}_{(3),\text{eff}} = -\mu_B\,(\mathbf{H}\cdot\mathbf{n})\sum_{\langle ijk\rangle} j_\Phi\,\tilde{\mathbf{S}}_i\cdot(\tilde{\mathbf{S}}_j \times \tilde{\mathbf{S}}_k). \tag{S26}$$

in terms of the unitless plaquette parameter:

$$j_\Phi = -\frac{1}{S}\frac{q}{\hbar}\frac{A_\text{triangle}}{\Phi}\,\tilde{J}_\chi^{ijk}, \tag{S27}$$

which is defined for each closed triangle plaquette $\langle ijk\rangle$. For the triangle pictured in Supplementary Figure 4, we estimated $j_\Phi \approx 0.039$. As noted in the main text, the operator $\tilde{\mathbf{S}}_i\cdot(\tilde{\mathbf{S}}_j\times\tilde{\mathbf{S}}_k)$ is isotropic, but these interactions provide explicit contributions to the magnetic torque through the appearance of $(\mathbf{H}\cdot\mathbf{n})$ in the coupling.

8

Finally, we have also considered higher order 4-spin ring-exchange couplings:

$$\mathcal{H}_{(4),\text{eff}} = \frac{1}{S^2} \sum_{\langle ijkl \rangle} \tilde{K}_{ijkl} (\tilde{\mathbf{S}}_i \cdot \tilde{\mathbf{S}}_j)(\tilde{\mathbf{S}}_k \cdot \tilde{\mathbf{S}}_l). \tag{S28}$$

The distinct four-site plaquette are labelled according to a horizontal ($K_h$), vertical ($K_v$) and diagonal ($K_d$) interaction (shown in Supplementary Figure 2c). In previous works in which the ring-exchange terms have been considered, the approximation has been typically taken that $K_h = K_v = K_d$ and $K' = \frac{J'}{J} K$[6,7]. However, these relations are not enforced by symmetries. Interestingly, as shown in Supplementary Table 1, we find that such relations do not hold when considering the full electronic structure of the dimers. While the effects of such terms on the ground state are a matter of intense study, the isotropic ring-exchange terms do not explicitly contribute to the torque, and therefore may be the subject of future studies.

## SUPPLEMENTARY NOTE 2: GENERALIZED TORQUE EXPRESSIONS

As mentioned in the main text, in general the magnetic torque is the derivative of the energy $E = \langle \mathcal{H} \rangle$ with respect to a reference angle $\theta$:

$$\tau = \frac{d\langle \mathcal{H} \rangle}{d\theta} \tag{S29}$$

This expression holds strictly in the $T \to 0$ limit, as entropic contributions are omitted for simplicity. After employing the site-dependent rotations described in Supplementary Note 1, the only terms in the Hamiltonian contributing to the magnetic torque are the uniform and staggered Zeeman terms and chiral 3-spin interactions. Here, we show the derivation of the bulk torque contribution for the uniform Zeeman term explicitly.

In terms of the uniform $g$-tensor $\tilde{\mathbb{G}}_u$ and laboratory field $\mathbf{H}$, the uniform Zeeman Hamiltonian is:

$$\mathcal{H}_{\text{Zee}} = -\mu_B \mathbf{H} \cdot \tilde{\mathbb{G}}_u \cdot \left( \sum_i \tilde{\mathbf{S}}_i \right). \tag{S30}$$

In general, we assume that $\langle \sum_i \tilde{\mathbf{S}}_i \rangle = 0$ at zero field. In the presence of a finite field, there are several subtleties that arise due to anisotropy in $\tilde{\mathbb{G}}_u$. For example, the Zeeman energy is minimized when the spins $\tilde{\mathbf{S}}_i$ are parallel to the effective field given by:

$$\mathbf{H}_{\text{eff},u} = \tilde{\mathbb{G}}_u^{\text{T}} \cdot \mathbf{H}, \tag{S31}$$

where $\mathbb{G}^\mathrm{T}$ denotes the transpose of $\mathbb{G}$. As a result, we may write:

$$\left\langle \sum_i \tilde{\mathbf{S}}_i \right\rangle = \chi_u \, \mathbf{H}_{\mathrm{eff},u}, \tag{S32}$$

in terms of a general susceptibility $\chi_u$. Since the Hamiltonian governing the response of the spins $\tilde{\mathbf{S}}_i$ is otherwise isotropic, and we consider a regime with no spontaneously broken symmetry, the susceptibility $\chi_u$ is isotropic with respect to the effective field, and therefore depends only on the magnitude $|\mathbf{H}_{\mathrm{eff},u}|$. We therefore assume that the susceptibility scales as a power law in terms of the magnitude of the effective field:

$$\chi_u = \tilde{\chi}_{0,u} \, |\mathbf{H}_{\mathrm{eff},u}|^{-\zeta_u} \tag{S33}$$

Combining these expressions, the uniform Zeeman energy is given by:

$$\langle \mathcal{H}_{\mathrm{Zee,eff},u} \rangle = -\mu_B \, \mathbf{H} \cdot \tilde{\mathbb{G}}_u \cdot \left\langle \sum_i \tilde{\mathbf{S}}_i \right\rangle \tag{S34}$$

$$= -\mu_B \, \tilde{\chi}_{0,u} \, |\mathbf{H}_{\mathrm{eff},u}|^{-\zeta_u} \left( \mathbf{H} \cdot \tilde{\mathbb{G}}_u \cdot \tilde{\mathbb{G}}_u^\mathrm{T} \cdot \mathbf{H} \right) \tag{S35}$$

$$= -\mu_B \, \tilde{\chi}_{0,u} \, |\mathbf{H}_{\mathrm{eff},u}|^{2-\zeta_u} \tag{S36}$$

To compute the torque, we need to take the angular derivative of this expression. The $\theta$-dependence arises from the norm of the effective field. For the derivative of the norm we use the following relation:

$$\frac{\mathrm{d}}{\mathrm{d}\theta} |\mathbf{H}_{\mathrm{eff},u}(\theta)|^\alpha = \alpha |\mathbf{H}_{\mathrm{eff},u}|^{\alpha-2} \frac{\mathrm{d}\mathbf{H}_{\mathrm{eff},u}}{\mathrm{d}\theta} \cdot \mathbf{H}_{\mathrm{eff},u}. \tag{S37}$$

The torque we then obtain is given by:

$$\tau_u(\theta) = \frac{-\mu_B (2-\zeta_u)\tilde{\chi}_{0,u} H^{2-\zeta_u}}{|\mathbb{G}_u^\mathrm{T} \cdot \mathbf{h}|^{\zeta_u}} \left( \frac{\mathrm{d}\mathbf{h}}{\mathrm{d}\theta} \cdot \tilde{\mathbb{G}}_u \cdot \tilde{\mathbb{G}}_u^\mathrm{T} \cdot \mathbf{h} \right). \tag{S38}$$

where $\mathbf{h}$ is a unit vector in the direction of the laboratory field $\mathbf{H}$, and $H$ is the magnitude of the laboratory field $H = |\mathbf{H}|$. In the main text, this expression is simplified by separating the $H$ and $\theta$ dependencies:

$$\frac{\tau(\theta)}{H^2} = \tilde{\chi}_u(H) f_u(\theta) \tag{S39}$$

$$\tilde{\chi}_u(H) = \mu_B (2-\zeta_u)\tilde{\chi}_{0,u} H^{-\zeta_u} \tag{S40}$$

$$f_u(\theta) = -\frac{1}{|\tilde{\mathbb{G}}_u^\mathrm{T} \cdot \mathbf{h}|^{\zeta_u}} \left( \frac{\mathrm{d}\mathbf{h}}{\mathrm{d}\theta} \cdot \tilde{\mathbb{G}}_u \cdot \tilde{\mathbb{G}}_u^\mathrm{T} \cdot \mathbf{h} \right) \tag{S41}$$

Analogous expressions follow for the staggered and chiral contributions to the torque.

It is useful to see that these expressions reproduce the conventional $\sin 2\theta$ dependence in the case when $\zeta = 0$. To show this, we consider the torque in the $a - c^*$ plane, with $\theta$ being the angle between $\mathbf{H}$ and $a$ within this plane. In this case, the corresponding torque is:

$$\tau_{a\text{-}c^*}(\theta) = -\tilde{\chi}_u \left[ \frac{\mathrm{d}}{\mathrm{d}\theta} \begin{pmatrix} H\cos\theta \\ 0 \\ H\sin\theta \end{pmatrix} \right] \cdot \tilde{\mathbb{G}}_u \cdot \tilde{\mathbb{G}}_u^{\mathrm{T}} \cdot \begin{pmatrix} H\cos\theta \\ 0 \\ H\sin\theta \end{pmatrix} \tag{S42}$$

$$= \tilde{\chi}_u H^2 \begin{pmatrix} \sin\theta \\ 0 \\ -\cos\theta \end{pmatrix} \cdot \tilde{\mathbb{G}}_u \cdot \tilde{\mathbb{G}}_u^{\mathrm{T}} \cdot \begin{pmatrix} \cos\theta \\ 0 \\ \sin\theta \end{pmatrix} \tag{S43}$$

Assuming the $g$-tensor is diagonal in the $a - c^*$ coordinates, this gives:

$$\frac{\tau_{a\text{-}c^*}(\theta)}{H^2} = \mu_B \, \tilde{\chi}_{0,u} \, (g_{aa}^2 - g_{c^*c^*}^2) \sin(2\theta) \tag{S44}$$

## SUPPLEMENTARY NOTE 3: IMPURITY SCALING

In this supplementary note, we present the derivation of the approximate scaling expressions for the impurity contributions to the magnetic torque discussed in the main text. Generically, the coupling of such "orphan spins" to the external field $\mathbf{H}$ is governed by the Zeeman Hamiltonian:

$$\mathcal{H}_{\text{Zee,I}} = -\mu_B \sum_m \mathbf{H} \cdot \tilde{\mathbb{G}}_{\mathrm{I},m} \cdot \tilde{\mathbf{S}}_{\mathrm{I},m} \tag{S45}$$

where $\tilde{\mathbb{G}}_{\mathrm{I},m}$ is the effective impurity $g$-tensor.

The response of the randomly coupled orphan spins can be understood with reference to the "strong disorder renormalization group" (SDRG) approach[8–13] to studying problems with quenched disorder. In the application to spin systems, the central quantity is the distribution of exchange couplings $\rho(J)$. An initial energy scale $\Omega$ is set by the strongest interaction within the network, which couples impurity spins $\tilde{\mathbf{S}}_{\mathrm{I},1}$, and $\tilde{\mathbf{S}}_{\mathrm{I},2}$. The relative degrees of freedom associated with these impurity spins $\tilde{\mathbf{S}}_{\mathrm{I},1}$ and $\tilde{\mathbf{S}}_{\mathrm{I},2}$ are then integrated out, yielding a new "cluster" with total effective spin $S_{\text{eff}} = |\tilde{\mathbf{S}}_{\mathrm{I},1} \mp \tilde{\mathbf{S}}_{\mathrm{I},2}|$, depending on the sign of $J_{12}$. This process modifies the effective interactions, yielding a new distribution

11

$\rho_\Omega(J)$ of interactions between clusters that is dependent on the energy scale $\Omega$. As $\Omega$ is successively lowered, the effective interactions between remaining spin clusters tend towards a fixed point power law distribution $\rho_\Omega(J) \sim J^{d/z-1}$, which gives rise to power law behaviour in relevant physical observables. Here, $d$ is the effective dimension, and $z$ is the dynamical critical exponent.

In order to derive approximate expressions for response in this scaling regime, it is useful to recast the summation over impurity spins as a summation over the clusters $C$.

$$\sum_m \langle \tilde{\mathbf{S}}_{I,m} \rangle = \sum_C^{N_C} \sum_{m \in C}^{n_C} \langle \tilde{\mathbf{S}}_{I,m} \rangle. \tag{S46}$$

where $N_C(\Omega)$ denotes the total number of such clusters and $n_C$ denotes the number of spins within the cluster $C$. As impurity spins become successively coupled at lower energies, $N_C$ decreases as $\Omega$ is lowered. At each energy scale $\Omega$, it is assumed that the distribution $\rho_\Omega(J)$ is sufficiently broad, that each impurity cluster is approximated as an independent "spin", with effective moment size given by $S_{C,\text{eff}}$. As a result, each cluster is described by a thermodynamic partition function:

$$\mathcal{Z}_C = \frac{\sinh\left((2S_{C,\text{eff}}+1)\frac{\mu_B|\tilde{\mathbb{G}}_I \cdot \mathbf{H}|}{2k_B T}\right)}{\sinh\left(\frac{\mu_B|\tilde{\mathbb{G}}_I \cdot \mathbf{H}|}{2k_B T}\right)} \tag{S47}$$

The contribution of each cluster to the torque is evaluated by taking the angular derivative of the free energy $G_C = -k_B T \ln \mathcal{Z}_C$, such that:

$$\tau = \sum_C^{N_C} \frac{dG_C}{d\theta} \tag{S48}$$

This yields:

$$\frac{\tau}{H^2} = \frac{\mu_B}{H} g(\theta) \sum_C^{N_C} S_{C,\text{eff}} \, \mathcal{B}\left(S_{C,\text{eff}}, \frac{\mu_B|\tilde{\mathbb{G}}_I^T \cdot \mathbf{H}|}{k_B T}\right) \tag{S49}$$

$$g(\theta) = -\left(\frac{d\mathbf{h}^T}{d\theta} \cdot \frac{\tilde{\mathbb{G}}_I \cdot \tilde{\mathbb{G}}_I^T}{|\tilde{\mathbb{G}}_I \cdot \mathbf{h}|} \cdot \mathbf{h}\right) \tag{S50}$$

In order to simplify this expression further, we introduce the cluster average moment:

$$S_{\text{eff}}^{\text{avg}} = \frac{1}{N_C} \sum_C^{N_C} S_{C,\text{eff}}, \tag{S51}$$

12

and approximate the cluster sum by:

$$\sum_C^{N_C} S_{C,\text{eff}} \mathcal{B}(S_{C,\text{eff}}, x) \approx N_C S_{\text{eff}}^{\text{avg}} \mathcal{B}(S_{\text{eff}}^{\text{avg}}, x). \tag{S52}$$

Similarly, we identify the energy scale with the largest of either the thermal energy or typical Zeeman energy of a cluster:

$$\Omega = \max(k_B T, \mu_B S_{\text{eff}}^{\text{avg}} |\tilde{\mathbb{G}}_\text{I}^\text{T} \cdot \mathbf{H}|). \tag{S53}$$

These expressions lead to the impurity torque expression given in the main text by Eq. (14). In practice, for the purpose of plotting, we use a soft maximum approximation, $\max(A, B) \approx (A^p + B^p)^{(1/p)}$ with $p = 2$. When plotted over several orders of $A/B$, the resulting functions are largely insensitive to the choice of $p$.

In order to evaluate the torque expression, the specific scaling of $N_C$ and $S_{\text{eff}}^{\text{avg}}$ with $\Omega$ is required. This depends on the nature of the disordered fixed point[10,11]. At any given energy scale, the number of independent clusters scales as $N_C(\Omega) \sim \int_0^\Omega \rho_\Omega(J) dJ \sim \Omega^{d/z}$, which is an increasing function of $\Omega$. For purely antiferromagnetic and unfrustrated interactions, the pairs of spins integrated out at any given energy scale would always form $S = 0$ singlets. As a result, no clusters of large moment would be formed as the energy is lowered, and $S_{\text{eff}}^{\text{avg}}$ would remain fixed. In contrast, in the presence of ferromagnetic or frustrated interactions[10,13], the average cluster moment must grow as $\Omega$ is successively lowered, scaling as $S_{\text{eff}}^{\text{avg}}(\Omega) \sim \Omega^{-\kappa}$ for some exponent $\kappa$. For purely ferromagnetic interactions, the average cluster moment would be directly proportional to the cluster size $S_{\text{eff}}^{\text{avg}} \propto N_C^{-1}$. For interactions with mixed signs, or frustrated antiferromagnetic couplings, the cluster moments increase more slowly $S_{\text{eff}}^{\text{avg}} \propto N_C^{-1/2}$, following the random walk argument of Ref. 10 and 13. Therefore, for $\kappa$-Cu and other frustrated systems, this suggests $d/z = 2\kappa$ is applicable:

$$N_C(\Omega) = N_0 \Omega^{2\kappa} \tag{S54}$$

$$S_{\text{eff}}^{\text{avg}}(\Omega) = S_0 \Omega^{-\kappa} \tag{S55}$$

for some constants $S_0$ and $N_0$.

In the low temperature or high-field limit $k_B T \ll \mu_B S_{\text{eff}}^{\text{avg}} |\tilde{\mathbb{G}}_\text{I}^\text{T} \cdot \mathbf{H}|$, solving Eq. (S54) and (S56) leads to:

$$\Omega \approx (\mu_B S_0 |\tilde{\mathbb{G}}_\text{I}^\text{T} \cdot \mathbf{H}|)^{\frac{1}{1+\kappa}}. \tag{S56}$$

In this limit, all cluster moments should be saturated, leading to the relation for the magnetization $m \sim N_C S_{\text{eff}}^{\text{avg}} \propto H^{-\zeta_I+1} \propto H^{\kappa/(1+\kappa)}$. Thus, the relation between the exponents is given by:

$$\zeta_I = \frac{1}{1+\kappa} = \frac{2z}{2z+d} \leq 1. \tag{S57}$$

Including prefactors, the reduced torque susceptibility follows from the definition of the torque as $\tau = \mu_B \, d(\mathbf{H} \cdot \mathbf{m}_I)/d\theta$, together with the above expressions:

$$\tilde{\chi}_I^H(H) = \tilde{\chi}_{0,I} \, H^{-\zeta_I} \tag{S58}$$

$$\tilde{\chi}_{0,I}^H = (2-\zeta_I)\frac{N_0\mu_B^2 S_0^2}{(\mu_B S_0)^{\zeta_I}} \tag{S59}$$

In the high temperature limit $k_B T \gg \mu_B S_{\text{eff}}^{\text{avg}}|\tilde{\mathbb{G}}_I^T \cdot \mathbf{H}|$ the energy scale is according to Eq. (S54):

$$\Omega \approx k_B T. \tag{S60}$$

With the scaling relations of $N_C$ (S55) and of $S_{\text{eff}}^{\text{avg}}$ (S56) and the relation between the exponent $\kappa$ and $\zeta_I$ (S58) the temperature dependence of the torque susceptibility, as given in Eq. (20) in the main text, follows from the expansion of the Brillouin function as $\mathcal{B}(S,x) = \frac{1}{3}(S+1)x$ for small $x$:

$$\tilde{\chi}_I^T(T) \approx \tilde{\chi}_{0,I}^T \left(\frac{1}{k_B T} + S_0^{-1}(k_B T)^{\frac{1}{\zeta_I-2}}\right) \tag{S61}$$

$$\tilde{\chi}_{0,I}^T = \frac{2}{3} N_0 \mu_B^2 S_0^2 \tag{S62}$$

Of particular note, this expression (for suitable constants) reproduces the lowest two orders in the expressions for the susceptibiltiy derived in Ref. 14 for random 1D chains.

Regarding the NMR linewidth, as discussed in Ref. 15–17, the orphan spin impurities contribute through the staggered moment induced in the surrounding bulk around each impurity. As a result, a nuclear spin at site $i$ in the bulk experiences a different effective local field, which is given by $\mathbf{H}_i = \mathbf{H} + \tilde{\mathbf{S}}_i \cdot \tilde{\mathbb{A}}_i$, in terms of the local hyperfine coupling tensor $\tilde{\mathbb{A}}$. We assume that the impurity-induced local magnetization is given by $\langle \tilde{\mathbf{S}}_i \rangle = a_i \langle \tilde{\mathbf{S}}_{I,m} \rangle$, where $m$ labels the impurity closest to the dimer site $i$. The constants $a_i$ are determined, for example, by the distance between $i$ and $m$. Thus, finite impurity moments will lead to a distribution of local fields, which then broadens the NMR lines according to the specific distribution of

$a_i$ and $\langle \tilde{\mathbf{S}}_{I,m} \rangle$. An important observation is that the magnitude of this broadening depends explicitly only on local quantities, rather than the cluster averages appearing in the total impurity torque. This leads to a different scaling of the NMR linewidth $\nu$ with field and temperature. However, within a given cluster, the impurity moments are assumed to remain perfectly correlated. As a result, the external field is able to orient the local impurity spins only through coupling to the total moments of the clusters. These observations can be summarized by the approximation:

$$\nu \propto \frac{1}{\sqrt{N}} \left[ \sum_m^N \langle \tilde{\mathbf{S}}_{I,m} \rangle \cdot \langle \tilde{\mathbf{S}}_{I,m} \rangle \right]^{\frac{1}{2}} \tag{S63}$$

where the NMR linewidth scales as the root-mean-square impurity magnetization. Here, $N$ is the total number of impurities. As before, we recast the summation in terms of clusters $C$:

$$\sum_m^N \langle \tilde{\mathbf{S}}_{I,m} \rangle \cdot \langle \tilde{\mathbf{S}}_{I,m} \rangle = \sum_C^{N_C} \sum_{m \in C}^{n_c} \langle \tilde{\mathbf{S}}_{I,m} \rangle \cdot \langle \tilde{\mathbf{S}}_{I,m} \rangle \tag{S64}$$

where $N_C$ gives the total number of clusters, and $n_C$ gives the number of original impurity spins within cluster $C$. In analogy with Eq. (S50), the contribution per cluster is approximated by:

$$\sum_{m \in C}^{n_c} \langle \tilde{\mathbf{S}}_{I,m} \rangle \cdot \langle \tilde{\mathbf{S}}_{I,m} \rangle \approx n_C \left[ \mathcal{B} \left( S_{C,\text{eff}}, \frac{\mu_B |\tilde{\mathbb{G}}_I^T \cdot \mathbf{H}|}{k_B T} \right) \right]^2 \tag{S65}$$

Note that $n_C$ appears as a prefactor here instead of $S_{C,\text{eff}}$ due to the fact that $\langle \tilde{\mathbf{S}}_{I,m} \rangle \cdot \langle \tilde{\mathbf{S}}_{I,m} \rangle > 0$. We then introduce that average cluster size as:

$$n_{\text{avg}} = \frac{1}{N_C} \sum_C^{N_C} n_C \tag{S66}$$

such that $n_{\text{avg}} N_C = N$. Finally, making the approximation:

$$\sum_C^{N_C} n_C \left[ \mathcal{B} \left( S_{C,\text{eff}}, x \right) \right]^2 \approx n_{\text{avg}} N_C \left[ \mathcal{B} \left( S_{\text{eff}}^{\text{avg}}, x \right) \right]^2 \tag{S67}$$

provides to the proposed expression:

$$\nu_I \approx \nu_0 \, \mathcal{B} \left( S_{\text{eff}}^{\text{avg}}, \frac{\mu_B |\tilde{\mathbb{G}}_I^T \cdot \mathbf{H}|}{k_B T} \right). \tag{S68}$$

where $S_\text{eff}^\text{avg}$ appears only in the argument of the Brillouin function.

Finally, it is important to consider experimentally relevant values for the nonuniversal exponents appearing in the scaling forms. As noted above, the torque response of the orphan spin defects is parameterized by the nonuniversal exponent $\zeta_\text{I}$, which is related to the low-energy distribution of effective interactions, $\rho_\Omega(J) \sim J^{2/\zeta_\text{I}-3}$. Although $\zeta_\text{I}$ is unknown *a priori*, practical considerations restrict $1 \lesssim z/d \leq \infty$, which corresponds to a narrow range $2/3 \lesssim \zeta_\text{I} \leq 1$. In general, $\zeta_\text{I}$ is likely to be sample dependent, and should tend to decrease with increasing frustration of the bulk interactions, and/or uniformity of the impurity distribution within the sample. Both such qualities lead to less singular fixed point distributions $\rho_\Omega(J)$. For example, the limit $\zeta \to 1$ (i.e. $z/d \to \infty$) corresponds to the *infinite randomness* limit, in which $\rho_\Omega(J)$ is maximally singular. At any given energy scale, the vast majority of spins remain essentially decoupled, leading to a Curie-like response $\chi \sim 1/T$, up to logarithmic corrections. Such a fixed point describes, for example, the random singlet phase (RSP)[8,9] for purely antiferromagnetic but random interactions in $d = 1$. In this case, the interactions are not frustrated, and strongly interacting pairs of spins always form singlets, such that large spin clusters do not form at low energies, suggesting $\kappa = 0$. In contrast, the opposite limit of a flat energy distribution, given by $\zeta \to 2/3$ (i.e. $z/d \to 1$), corresponds to a so-called "spin-glass" fixed point (SGFP)[10]. This fixed point has been found in $d = 2$ via the SDRG appraoch for both geometrically frustrated lattices with random but purely antiferromagnetic interactions, as well as bipartite lattices with mixed ferro- and antiferromagnetic couplings[10]. In both cases clusters with large $S_\text{eff}$ are generated at lower energies. Lying between these extremes are the so-called "large-spin" fixed points (LSFP), which have $0 < \kappa < 1/2$. For example, the inclusion of both ferro- and antiferromagnetic couplings in $d = 1$ leads to a LSFP[12–14] with $\kappa = 0.21$, $\zeta = 0.83$. Similarly, a LSFP also describes randomly site-diluted models in $d = 2$ with purely antiferromagnetic interactions, yielding $\kappa \sim 0.1-0.2$, depending on the degree of dilution. This corresponds to $\zeta \sim 0.8-0.9$.

In principle, the impurities in $\kappa$-Cu should correspond to a random $d = 2$ lattice with *both* site dilution and random ferro-/antiferromagnetic couplings. To the best of our knowledge, the appropriate exponents have not yet been studied for this case. However, it should be emphasized that a relatively large variation in $z/d$ leads to narrow range of susceptibility exponents $2/3 \lesssim \zeta \leq 1$. On this basis, we conclude that the experimental values of $\zeta_\text{exp} = 0.76 - 0.83$ observed for $\kappa$-Cu fall well within the range expected for impurity effects. The

16

observed variance $\Delta\zeta = 0.07$ corresponds to about 20% of the realistic range, which may be an indication of strong sample dependence.

## SUPPLEMENTARY REFERENCES


\end{document}